\documentclass[a4paper,fleqn,usenatbib]{mnras}

\usepackage{mathptmx}

\usepackage[T1]{fontenc}
\usepackage{ae,aecompl}

\usepackage{graphicx}	
\usepackage{amsmath}	
\usepackage{amssymb}	
\usepackage{units}
\usepackage{array}
\usepackage{relsize} 
\usepackage{soul}

\newcommand{\kms}{\, \mathrm{km} \, \mathrm{s}^{-1}}
\newcommand{\Teff}{T_\mathrm{eff}}
\newcommand{\vect}[1]{\boldsymbol{#1}}
\newcommand{\lv}{\ell_v}
\DeclareMathOperator*{\argmax}{arg\,max}

\title[Kinematical age method]{A method to estimate stellar ages from kinematical data}

\author[F. Almeida-Fernandes \& H. J. Rocha-Pinto]{
F. Almeida-Fernandes,
H. J. Rocha-Pinto
\\
Observat\'orio do Valongo, Universidade Federal do Rio de Janeiro -- UFRJ, Ladeira Pedro Ant\^onio 43, 20080-090 Rio de Janeiro, RJ, Brazil\\
}

\date{Accepted XXX. Received YYY; in original form ZZZ}

\pubyear{2016}

\begin{document}
\label{firstpage}
\pagerange{\pageref{firstpage}--\pageref{lastpage}}
\maketitle

\begin{abstract}
We present a method to build a probability density function (pdf) for the age of a star based on its peculiar velocities $U$, $V$ and $W$ and its orbital eccentricity. 
The sample used in this work comes from the Geneva-Copenhagen Survey (GCS) which contains both the spatial velocities, orbital eccentricities and isochronal ages for about $14\,000$ stars.
Using the GCS stars, we fitted the parameters that describe the relations between the distributions of kinematical properties and age. This parametrization allows us to obtain an age probability from the kinematical data. From this age pdf, we estimate an individual average age for the star using the most likely age and the expected age.
We have obtained the stellar age pdf for the age of $9\,102$ stars from the GCS and have shown that the distribution of individual ages derived from our method is in good agreement with the distribution of isochronal ages. We also observe a decline in the mean metallicity with our ages for stars younger than 7 Gyr, similar to the one observed for isochronal ages. This method can be useful for the estimation of rough stellar ages for those stars that fall in areas of the HR diagram where isochrones are tightly crowded. As an example of this method, we estimate the age of Trappist-1, which is a M8V star, obtaining the age of $t(UVW) = 12.50(+0.29-6.23)$ Gyr.

\end{abstract}

\begin{keywords}
stars: kinematics and dynamics -- solar neighbourhood
\end{keywords}


	
\section{Introduction}

Even though our knowledge about the Galaxy has remarkably grown in the last century, we are still struggling with fundamental questions regarding its formation, structure and evolution. Understanding the chemodynamical evolution of the Galaxy from its birth to the present day is challenging because we can only directly observe a single frame in time, from an otherwise ever--changing complex scenario.

One way to infer the properties of the Galaxy at different epochs, and then recover its chemodynamical evolution, is by using individual stellar ages. The chemical abundances of a star reflects the interstellar medium abundances at the time in which the star was formed, and its kinematical properties may contain information about the galactic structure at that particular epoch.

Individual stellar ages are also important for studies of single stars, as the whole physical structure of a star may be determined exclusively from its mass, chemical composition and age \citep{Vogt1926, Russell+1927}. Mass and chemical abundances can be directly measured but ages can only be inferred from observable properties that are known to change with time \citep{Soderblom2010}.

The task of estimating stellar ages has been addressed by several authors and lots of different methods are found in the literature. For instance, there are: (i) empirical methods, which uses a deterministic relation between a given parameter and the age of a star (first proposed by \citealp{Skumanich1972}); this is the case of gyrochronology \citep[e.g.][]{Barnes2003, Barnes2007, Mamajek+Hillenbrand2008, CollierCameron+2009}, decay of cromosferic activity \citep[e.g.][]{Soderblom+1991, Rocha-Pinto+Maciel1998, Rocha-Pinto+2000, Lyra+PortodeMello2005, Pace+Pasquini2004, Pace+2009, Pace2013, Zhao+2011}, lithium depletion \citep{Sestito&Randich2005, Jackson+Jeffries2014, Carlos+2016} and ``magnetochronology'' (proposed by \citealp{Vidotto+2014}). (ii) model dependent methods, which are based on the comparison between measurable physical quantities and the ones expected from stellar structure models that use age as one of its parameters. Isochrone fitting \citep{Edvardsson+1993, Pont+Eyer2004, Nordstrom+2004, Jorgensen+Lindegren2005, Silaj+Landstreet2014, Maxted+2015} and asteroseismology \citep{Cunha+2007, Vauclair2009, Metcalfe+2010, Silva-Aguirre+2017} are classified in this category. (iii) semi-fundamental methods, those that are based on well known fundamental physics and employ only few assumptions; these are the cases for the method of cluster expansion \citep[e.g.][]{Makarov2007}, and nucleocosmocronology, known to predict unreliable ages \citep{Ludwig+2010}. (iv) Statistical methods, which uses statistical relations, like the age-metallicity relation (AMR) and the age--velocity dispersion relation (AVR), between a given property and the age.  These relations have not been much explored in the literature as a direct tool to estimate stellar ages. Some few examples of its usage are found in \citet{Lachaume+1999} and \citet{Maciel+2011} (for the AVR) and \citet{Spina+2016} (for the AMR, specially [Y/Mg]$\times$age and [Y/Al]$\times$age). 

\citet{Maciel+2011} developed a kinematical method that is based on the difference between the actual stellar rotational velocity, $\Theta$, and the one expected from the rotational curve. This difference is attributed to occasional disturbances on the stellar orbit and, therefore, stars that show greater differences must be older because they have participated in more collisions. In their work, \citet{Maciel+2011} adopt a deterministic relation between a given kinematical parameter and the stellar age. But the fact that stars of all ages may possess almost any velocity at all (e.g. the low velocities old stars, and the rare high velocities young stars) goes against their method. What is for sure related to the age is the probability of a star having a given velocity, because the velocity distribution changes based on the average age of the stellar sample. \st{} \citep{Nordstrom+2004, Casagrande+2011}.

It is known that the velocity dispersion of stars increases with the average age of the stellar groups \citep{Wielen1974, Wielen1977, Nordstrom+2004, Koval+2009, Casagrande+2011, Gontcharov2012, Martig+2014}. This relation has also been observed in external galaxies \citep{Beasley+2015, Dorman+2015}. The cause for this effect may be due to (i) stars being born dynamically hotter in the past or (ii) to stars being dynamically heated through time. The former can be explained by the disk being more gas-rich and more turbulent in the past \citep[e.g.][]{Bournaud+2009}, which is supported by observations of high-redshift galaxies \citep{Forster+2009}. Several causes have been proposed to explain the latter mechanism and are supported by cosmological simulations \citep[e.g.][]{House+2011}: heating caused by giant molecular clouds \citep[e.g.][]{Lacey1984, Hanninen+Flynn2002}, interaction with non-axissymetric galactic structures as transient and recurring spiral arms \citep[e.g.][]{Carlberg+Sellwood1985, Martinez-Medina+2015} or the bar \citep[e.g.][]{Saha+2010, Grand+2016}, and also interactions with satelite galaxies \citep[e.g.][]{Velazquez+White1999}.

Differently from \citet{Maciel+2011}, who obtained ages from a deterministic relation, we make use of the whole velocity distribution function and its well studied relation with age, to derive a new statistical age dating method, achieved through a Bayesian approach.

It is important to notice, that not all authors agree that the velocity dispersion parametrization is adequate to represent the velocity distribution in the $UV$ plane \citep[e.g.][]{Seabroke+Gilmore2007}. The reason for this is the existence of several substructures with typical sizes of $\approx 10 \, \kms$ in this plane. The mapping and study of the origins of these substructures are a very hot topic in the present days \citep{Dehnen1998, Famaey+2007, Bovy+Hogg2010, Bobylev+Bajkova2016}. For instance, it is not clear if these substructures are only local, or if they cover the whole disk \citep{Kushniruk+2017}. If their origin is dynamical, they would cover a large spread of ages, or would be concentrate within a single age if they originate from the disruption of stellar clusters \citep{Famaey+2008}. Considering that their origins and relation with age are still not well understood, and that modelling all these substructures in the velocity distributions would result in a very large number of free parameters, we still choose to work with the age-velocity dispersion. However, our model allows for an age-dependent correlation between the $U$ and $V$ velocities (described by the vertex deviation, $\lv$), which takes into account the existence of these groups as a first order approximation.

We work with F and G dwarfs because they have a long life expectancy. Therefore, their properties reflect properties of the disk at many different epochs. These stars also do not live too long to cause their internal changes to be insufficient to result in significant variations on observational properties that are used to derive their ages. Since orbital diffusion is not expected to depend on the stellar masses, the methods we derived for F and G stars are also likely to be valid for K and M dwarfs, which generally cannot have ages estimated by isochrones. As these stars have a lifetime that is greater than the age of the disk, they may contain key information about its chemodynamical state in different epochs and they are also numerous enough for us to work with averaged properties instead of individual stars that may be somehow peculiar. In Section \ref{sec:trappist1} we exemplify the flexibility of the method by calculating the age of Trappist-1, an M8V star for which other methods are not reliable.

We show how this Bayesian approach can be used to obtain a probability density function from which independent statistical ages may be estimated. We also show that from these estimated ages, important constraints on the evolution of the Galaxy may be obtained, as the age distribution and the age-metallicity relation.

The paper is structured as follows: Section \ref{sec:sample} presents the sample used to perform our analysis. In Section \ref{sec:methods} we develop three kinematical methods that can be used independently. The discussion of the method and some of our results are shown in Section \ref{sec:discussion}. Finally, Section \ref{sec:conclusions} presents our conclusions.
 
\section{Sample}
\label{sec:sample}

In order to calibrate the relations between the stellar kinematical parameters and the ages, we need a sample of stars that have known ages and velocities. The Geneva-Copenhagen Survey \citep{Nordstrom+2004}, an all-sky survey of F and G dwarf stars in the Solar Neighbourhood, is ideal for this task. It contains isochronal ages for most of the stars as well as unbiased kinematical information.

The first version of the catalog \citep{Nordstrom+2004} presented around $63\,000$ new radial velocities measurements. Together with published \textit{uvby$\beta$} photometry, Hipparcos parallaxes and Tycho-2 proper motions, this data allowed the authors to calculate, for most of the stars, their spatial velocities and also derive effective temperatures and metallicities from photometric calibrations. With these astrophysical parameters and theoretical isochrones from the Padova group \citep{Girardi+2000, Salasnich+2000} they applied the Bayesian method described by \citet{Jorgensen+Lindegren2005} to derive isochronal ages.

Since its first publication, there were three revisions of this survey. \citet{Holmberg+2007} improved the photometric calibrations for early F stars, \citet{Holmberg+2009} re-derived the astrophysical parameters implementing the revision of the Hipparcos parallaxes by \citet{vanLeeuwen2007}, and \citet{Casagrande+2011} revised the data using the effective temperature scale presented in \citet{Casagrande+2010} and also obtained new metallicity scales. From this improved data, \citet{Casagrande+2011} derived isochronal ages for the sample stars using both the Padova \citep{Bertelli+2008, Bertelli+2009} and BaSTI \citep{Pietrinferni+2004,Pietrinferni+2006} grids.

We built our sample from the data of \citet{Casagrande+2011} after applying the following selection criteria: (i) first, we removed all the stars for which the kinematical data was not complete; (ii) to avoid unknown binaries for which the derived astrophysical parameters would be unreliable, we discarded stars that had radial velocity determined from a single observation; (iii) as we are interested in fitting relations between kinematical parameters and age, we considered only the stars that could have its age derived by \citet{Casagrande+2011}; (iv) in order to work only with stars that have well defined ages, we have removed all those whose difference between the 84\% and 16\% percentiles and the median of the age probability density function exceeds 3 Gyr; (v) also to avoid unreliable ages, we have selected only the stars whose difference between the isochronal ages from the BaSTI and Padova grids are smaller than 1 Gyr. After applying all these selection criteria, our final sample consists of $9\,102$ F and G dwarf stars having complete kinematical data and well defined isochronal ages.

For simplicity, we consider further only the ages derived by the Padova grid. This will not affect our results because, by our sample definition, Padova and BaSTI ages cannot be significantly different. 

\section{Methods}
\label{sec:methods}

The relation between age and velocity dispersion means that the probability of a star having a given velocity depends on the stellar age. Through the Bayes theorem, this probability relation can be reversed and one can obtain a probability density function for the age of a star given their velocities.

In its most general form, the Bayes Theorem for the probability of a parameter $\theta$, including multi-dimensional data $\vect{d}$ and general background knowledge $I$ can be written as
\begin{equation}
\label{eq:bayes_general_form}
p(\theta|\vect{d},I) \propto p(\vect{d}|\theta,I) \, p(\theta|I) \,.
\end{equation} 

In the case the observational data consists of two observables (i.e. $d_1$ and $d_2$), the pdf for the parameter $\theta$ is given by \citep[eq. 48]{dAgostini2003}
\begin{equation}
\label{eq:bayes_general_two_obs}
p(\theta|d_1, d_2 ,I) \propto p(d_2|d_1,\theta,I) \, p(d_1|\theta,I) \, p(\theta|I) \,.
\end{equation}

When two observables are independent, it is also valid that
\begin{equation}
\label{eq:bayes_general_two_obs}
p(d_2|d_1,\theta,I) = p(d_2|\theta,I) \,.
\end{equation}

In this section, we show how these relations can be used to derive a probability density function for the age of a star from its measured spatial heliocentric velocities $U$, $V$, $W$ (named here as Method $UVW$). We also show how other kinematical properties, such as the eccentricity, can be used in the formalism, which is the case of Methods $eVW$ and $eUW$.

\subsection[Method $UVW$]{Method $\boldsymbol{UVW}$}
\label{sec:method_UVW}

In this method, the observational data consists of the spatial heliocentric velocity components $U$, $V$ and $W$ \footnote{In this work we define the $U$ axis as directed toward the galactic center.}. An important factor that must be taken into account is the existence of correlation between the $U$ and $V$ components, that might also depend on age \citep[see, for instance,][]{Rocha-Pinto+2004}. The correlations involving the $W$ component are usually smaller than the errors and therefore can be ignored \citep{Binney+Merrifield1998}.

The equations in this method are simplified by working with the components of the velocity ellipsoid, $v_1$, $v_2$ and $v_3$, instead of $U$, $V$ and $W$. This is the case because, by definition, they have no correlation between them. The $W$ and $v_3$ components are equivalent except for the displacement that is necessary to ensure zero mean for $v_3$, which is known to be caused by the Solar velocity with respect to the Local Standard of Rest ($W_{\sun}$). Calculating $v_1$ and $v_2$ then involves only the $U$ and $V$ components and can be described in terms of the vertex deviation ($\ell_v$). To calculate $v_1$ and $v_2$ one also needs to know the solar velocities $U_{\sun}$ and $V'_{\sun}$, with special care in the case of $V'_{\sun}$ because it may also depend on age. The transformation between $U$, $V$ and $W$ is then given by
\begin{subequations}
\label{eq:v1v2v3}
\begin{align}
v_1 &= (U+U_{\sun}) \, \cos{\lv} + (V+V'_{\sun}) \, \sin{\lv} \,, \\
v_2 &= -(U+U_{\sun}) \, \sin{\lv} + (V+V'_{\sun}) \, \cos{\lv} \,, \\
v_3 &= W+W_{\sun} \,,
\end{align}
\end{subequations}
where
\begin{equation}
\lv = \frac{1}{2}\arctan{\left( \frac{2\,\sigma_{UV}}{\sigma^2_U-\sigma^2_V} \right)}\,.
\end{equation}

\begin{figure}
\centering
\includegraphics[scale=1]{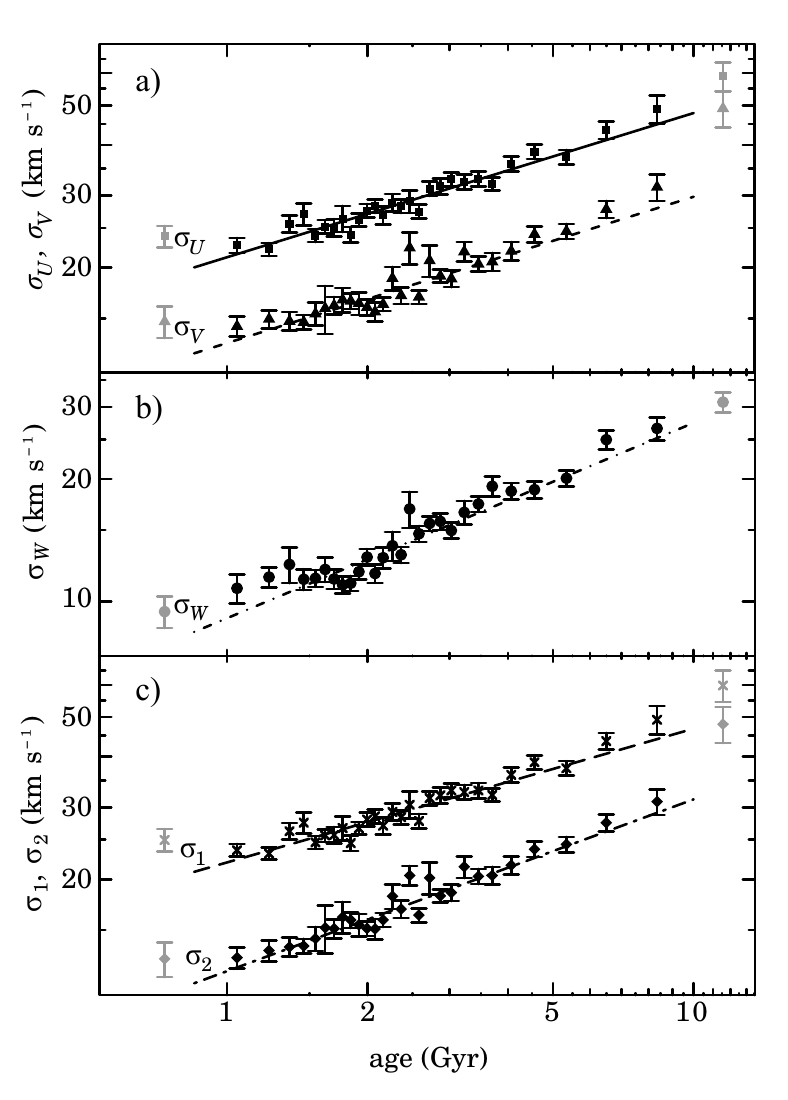}
\caption{Velocity dispersion as a function of age for the $U$ (squares, solid line), $V$ (triangles, dashed line) and $W$ (circles, dot-dashed line) components, and also for the principal components, $v_1$ (open squares, long-dashed line), $v_2$ (open triangles, double dashed line), for each of the 30 bins divided by age. Uncertainties were estimated by bootstrap re-sampling. For each component, the line represents the best fit as a function of age following a relation $\sigma_i = b_i \, t^{a_i}$. The first and last group were excluded from the fit for reasons explained in the text. The values fitted for the parameters $b_i$ and $a_i$ are shown in Table \ref{tab:fit_sigma}.}
\label{fig:sigma_fit}
\end{figure}

The transformation between the probability density functions is given in terms of the Jacobian of the coordinate transformation as $p(t|U,V,W) = J(v_1,v_2,v_3) \, p(t|v_1,v_2,v_3)$, where, through the remaining of the paper, $t$ denotes age. In this case this transformation is very simple for it involves only translations and rotations so the Jacobian is equal to unity. It then follows from Equation \ref{eq:bayes_general_form} (omitting the term denoting general background knowledge, $I$) that
\begin{equation}
p(t|U,V,W) \propto p(v_1|t) \, p(v_2|t) \, p(v_3|t) \, p(t) \,.
\end{equation}

The probability $p(t)$ corresponds to the prior age probability, before including the observed data. The only known information we are considering is that no star in the galaxy should be older than 14 Gyr. For all lower ages we consider a uniform probability distribution, so that $p(t)$ is given by
\begin{equation}
p(t) = \begin{cases}
	1, & \mathrm{if} \; 0 < t < 14 \; \textrm{Gyr.} \\
	0, & \mathrm{otherwise.}
\end{cases}
\end{equation}

To describe the probabilities $p(v_i|t)$ we approximate the distributions as Gaussians having dispersions that depends on age, $\sigma_i(t)$. This simplification is necessary because we need to work with distributions that can be described by few parameters, for which we can fit relations as a function of age from our data of $9\,102$ stars. A more realistic description of the stellar distribution, that also includes moving groups, is beyond the scope of this work. 

The parameters that are a function of age are the dispersions $\sigma_1(t)$, $\sigma_2(t)$, $\sigma_U(t)$, $\sigma_V(t)$ and $\sigma_W(t)$, and also the 
vertex deviation $\lv(t)$ and the $V$ component of the Solar motion $V'_{\sun}$. In order to fit the age dependencies of these parameters we use a procedure similar to the one used by \citet{Nordstrom+2004} for the original GCS sample. First, we have divided the sample in 30 bins according to the stellar ages, then, the aforementioned parameters were calculated for each of the bins and the uncertainties were estimated by bootstrap re-sampling. Figure \ref{fig:sigma_fit} shows the velocity dispersion for each component calculated for each one of the groups. As usually done in the literature, we have chosen to fit the relation between dispersion and age as single power laws in the form $\sigma_i = b_i \, t^{a_i}$, with $t$ in Gyr. In each case, we excluded the first and last bins, shown in grey, before fitting the relation. The reason for this is that the last bin may be contaminated with stars from the thick disk, biasing the dispersion to higher values, and the stars that belong to the first bin are too young and have not completed a significant amount of orbits so they may reflect the kinematical properties of not yet dissolved local structures. We present the values obtained for $a_i$ and $b_i$ for all components in Table \ref{tab:fit_sigma}. The results corroborate what has already been established from the GCS: there is an increase in the velocity dispersion for all components in all age ranges considered for the disk. 

\begin{table}
	\centering
	\caption{Parameters $b_i$ and $a_i$ that parametrize the relation between velocity dispersion and age, $\sigma_i = b_i \, t^{a_i}$, for each of the components $U$, $V$ and $W$ and for the principal components $v_1$ and $v_2$. Values obtained by \citet[GCSI]{Nordstrom+2004} and \citet[GCSIII]{Holmberg+2009} are shown for comparison.}
	\label{tab:fit_sigma}
	\begin{tabular}{ccccc} 
		\hline
		 & $b$ & $a$ & $a_{GCSI}$ & $a_{GCSIII}$ \\
		\hline
		$U$ & $21.2 \pm 1.0$ & $0.35 \pm 0.02$ & 0.31 & 0.39\\
		$V$ & $13.0 \pm 1.0$ & $0.36 \pm 0.02$ & 0.34 & 0.40\\
		$W$ & $9.1 \pm 1.0$ & $0.48 \pm 0.04$ & 0.47 & 0.50\\
		$v_1$ & $22.0 \pm 1.0$ & $0.33 \pm 0.02$ & & \\
		$v_2$ & $11.9 \pm 1.0$ & $0.42 \pm 0.02$ & & \\
		\hline
	\end{tabular}
\end{table}

\begin{figure}
\centering
\includegraphics[scale=0.9]{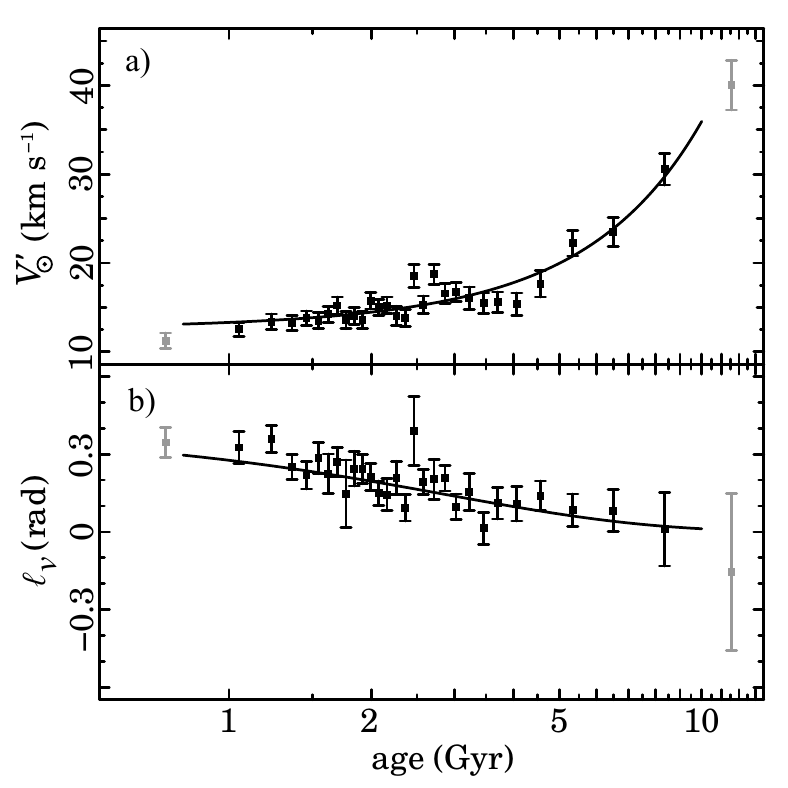}
\caption{Vertex deviation, $\lv$ (bottom), and $V'_{\sun}$ (top) calculated for each of the 30 groups dived by ages. Uncertainties were estimated by bootstrap re-sampling. The relations that describe this parameters as function of age (solid lines) are fitted as $V'_{\sun}(t) = 0.17 \, t^2 + 0.63 \, t + 12.5$ and $\lv(t) = 0.41 \, \exp(-0.37 \, t)$. For reasons explained in the text, the first and last groups were removed before fitting the relations.}
\label{fig:meanv_lv_fit}
\end{figure}

The relations between the remaining parameters, $\lv$ and $V'_{\sun}$, and age were found in a similar manner and are represented in Figure \ref{fig:meanv_lv_fit}. The expressions that best describe the behaviour of these parameters as a function of age are
\begin{align}
\lv(t) &= 0.41 \, \exp(-0.37 \, t) \\
V'_{\sun}(t) &= 0.17 \, t^2 + 0.63 \, t + 12.5 \, \mathrm{.}
\end{align} 
The increase found for $V'_{\sun}$ with age is expected and is caused by the asymmetric drift. 

There is still much discussion regarding the cause of the vertex deviation and its dependence on age. Here we find that the vertex deviation is maximum for the youngest stars, and tends to zero for the older stars. A possible cause for the vertex deviation are the presence of moving groups \citep{Dehnen1998}.

For instance, an investigation regarding the existence and origin of the main moving groups in the Solar neighbourhood was done by \citet{Famaey+2008}. The positions of the four moving groups found in his study (Hercules, Pleiades, Hyades and Sirius) are aligned with the necessary rotation we found for the principal component of the velocity elipsoid. The stars in these groups are mostly associated with young clusters of a few hundreds of Myr, which corroborates the decline of the vertex deviation with age.

\begin{figure*}
\centering
\includegraphics[scale=0.9]{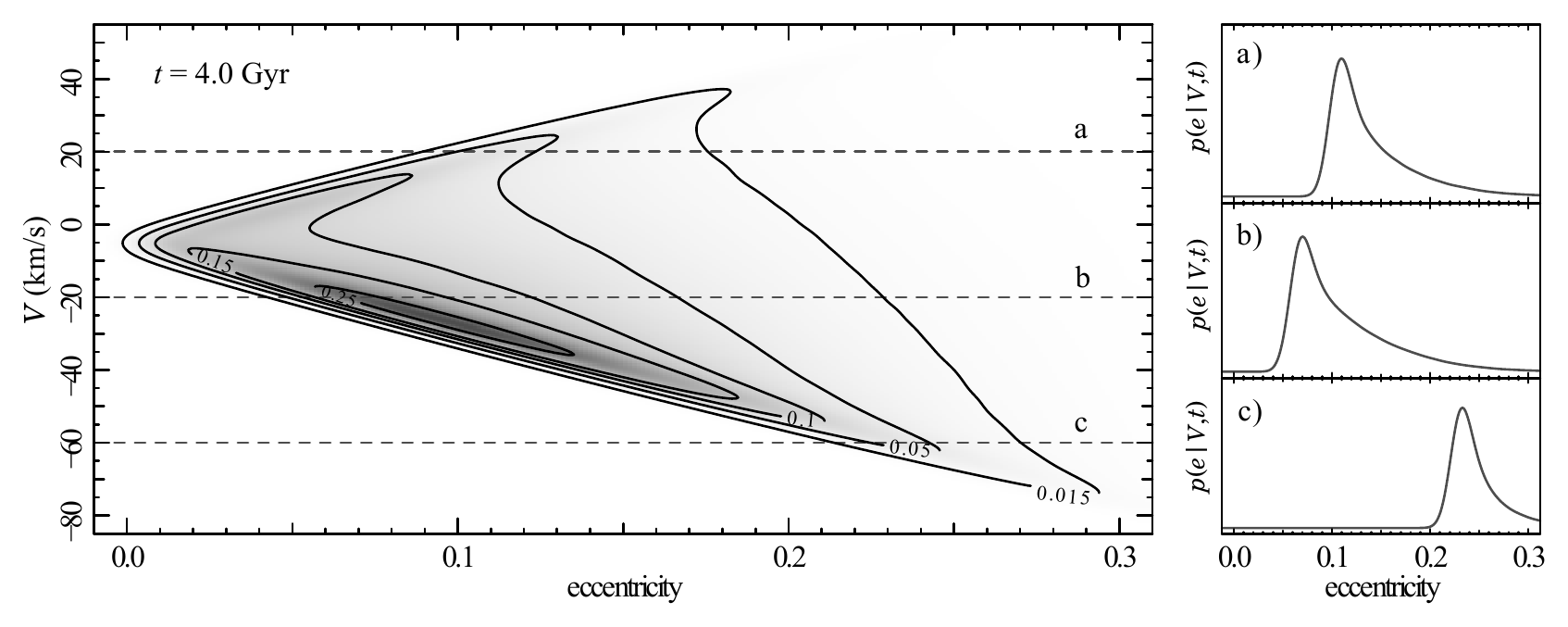}
\caption{(left) $eV$ plane density map built from a simulated sample of $2\,000\,000$ stars having $t = 4$ Gyr. The eccentricity was considered a function of $U$ and $V$, which were sampled from a distribution described by the parameters presented in Section \ref{sec:method_UVW}. (right) The probability density functions obtained from the density map for the eccentricity given the age (4 Gyr) and three different velocities $V = 20, -20, -60 \kms$, respectively (also represented on the left panel by dashed lines).}
\label{fig:dens_map}
\end{figure*}

The Solar velocity components are found from the averages of the velocities of stars in the Solar Neighbourhood. Our sample results in $U_{\sun} = 9.8 \pm 0.3 \kms$ and $W_{\sun} = 7.2 \pm 0.2 \kms$. As the average for the azimuthal component, $V$, has an age dependence, deriving this component of the Solar velocity is not as simple. If we define the $V_{\sun}$ component to be the velocity obtained from an idealized sample of stars with zero age, equivalent to $V'_{\sun}(t=0)$, we obtain $V_{\sun} = 12.5 \pm 0.9 \kms$. Table \ref{tab:sun_uvw} shows a comparison between this values and those obtained by other authors. As can be seen, the values obtained by different authors differ considerably, but the values calculated in this work are within the usual ranges. In Section \ref{sec:uncertainties-Solar-motion} we investigate how different input Solar Velocities affects the ages determinations.

\begin{table}
  \centering
	\caption{Comparisson between the values found for the Solar velocity components in this work, and the values obtained by \citet[F\&A14]{Francis+Anderson2014}, \citet[B\&B14]{Bobylev+Bajkova2014}, \citet[Co{\c s}+11]{Coskunoglu+2011}, \citet[Sch+10]{Schonrich+2010} and \citet[Kov+2009]{Koval+2009}.}
	\label{tab:sun_uvw}
	\begin{tabular}{lccc} 
		\hline
		Author & $u_{\sun} \kms$ & $v_{\sun} \kms$ & $w_{\sun} \kms$\\
		\hline
		This work & $9.8 \pm 0.3$ & $12.5 \pm 0.9$ & $7.2 \pm 0.2$\\
		F\&A14 & $14.1 \pm 1.1$ & $14.6 \pm 0.4$ & $6.9 \pm 0.1$\\
		B\&B14 & $6.0 \pm 0.5$ & $10.6 \pm 0.8$ & $6.5 \pm 0.3$\\
		Co{\c s}+11 & $8.83 \pm 0.24$ & $14.19 \pm 0.34$ & $6.57 \pm 0.21$\\[+0.03in]
		Sch+10 & $11.1^{+0.69}_{-0.75}$ & $12.24^{+0.47}_{-0.47}$ & $7.25^{+0.37}_{-0.36}$\\[0.03in]
		Kov+09 & $5.1 \pm 0.4$ & $7.9 \pm 0.5$ & $7.7 \pm 0.2$\\
		\hline
	\end{tabular}
\end{table}

We then build the age pdf calculating the probability $p(t|U,V,W)$ for different ages. As we assume a Gaussian distribution, this probability is given in terms of the adjusted parameters as\footnote{The symbols $\sigma_1$ and $\sigma_2$ represent the velocity dispersions for the components $v_1$ and $v_2$. It should also be noted that Equation \ref{eq:v1v2v3} implies that $\sigma_3$ = $\sigma_W$.}

{\medmuskip=0mu
\thinmuskip=0mu
\thickmuskip=0mu
\begin{equation}
\label{eq:prob_t_method1}
p(t|U,V,W) \, \propto  \mathlarger{\mathlarger{\prod_{\mathsmaller{\mathsmaller{i = 1, 2, 3}}}}} \left[ \frac{1}{(2\,\pi)^{\nicefrac{1}{2}} \, \sigma_i(t)} \exp{ \left(-  \frac{v_i^2}{2 \sigma_i(t)^2}\right)} \right] \,,
\end{equation}}
where $v_1$, $v_2$ and $v_3$ are obtained from Equation \ref{eq:v1v2v3}.

\subsection[Method $eVW$]{Method $\boldsymbol{eVW}$}
\label{sec:method_eVW}

One can also use other kinematical parameters instead of the stellar velocity components, as long as the distribution for these parameters are known and varies with age. In this work, we show how this can be done including the orbital eccentricity. We call it Method $eVW$. Now the probability is obtained from the eccentricity and the velocities $V$ and $W$. In this case, these variables are no longer independent, since the eccentricity correlates with the $UV$ velocities, so the Bayes equation gives us
\begin{equation}
p(t|e,V,W) = p(e|V,t) \, p(V|t) \, p(W|t) \, p(t) \,.
\end{equation}

As we did for the Method $UVW$, we use a uniform $p(t)$ for ages between 0 and 14 Gyr and approximate the velocity distribution as Gaussians. The relation between the velocity distribution parameters and age were already obtained in Section \ref{sec:method_UVW} and all that is left to know is the probability $p(e|V,t)$.

In this work, we obtain the probability $p(e|V,t)$ from density maps in the $eV$ plane for different ages. To create the density maps, we generate random pairs of values ($U$, $V$) following the distribution described by the parameters obtained in Section \ref{sec:method_UVW} for a given age $t$. Then, we estimate the eccentricity for the simulated $U$ and $V$ values and build a density map using a 2d kernel density estimator method. The approximate expression used to estimate the eccentricities was
{\medmuskip=0mu
\thinmuskip=0mu
\thickmuskip=0mu
\begin{equation}
e = 2.98 \cdot 10^{-3} \, \left( 155 + 20.0\,U + 19.6\,V + U^2 + 1.95\,V^2 \right)^{\nicefrac{1}{2}} - 7.23 \cdot 10^{-4} \,,
\label{eq:ecc}
\end{equation}}
which was obtained with the automated model building software Eureqa \citep{Schmidt+Lipson2009}. This software iteratively tries a combination of polynomial, linear, exponential, logarithmic, trignometric and power functions, penalizing complexity, to find the best relation to describe a given parameter as a function of multiple others in a data set. In this case, we obtained the expression for the eccentricity as a function of the velocities $U$, $V$ and $W$ using the data set of \citet{Casagrande+2011}, therefore, this eccentricity corresponds to the one obtained when using the same galactic potential these authors have used in their paper.

This simple expression for the eccentricity eliminates the need for orbital integration, simplifying the process of building the density maps. Compared to the eccentricity calculated by \citet{Casagrande+2011} for the GCS stars, we found that, for 98\% of them, the eccentricity calculated by this expression differs from the one given in the catalog by less then 0.01, which we consider to be good enough for our purposes.

To exemplify how drastically the probability $p(e|V,t)$ may change for different values of $V$ and same age $t$ we present in Figure \ref{fig:dens_map} the density map in the $eV$ plane for the age of 4 Gyr. Also in Figure \ref{fig:dens_map}, the probability functions $p(e|V,t)$,  for assigned values of $V$, are represented. 

Considering the known values of $e$, $V$ and $W$ for a star, the probabilities $p(e|V,t)$, $p(V,t)$ and $p(W,t)$ are then calculated for different ages $t$ and the results are used to build the probability density function for the stellar age. While $p(e|V,t)$ is obtained through the density maps, $p(V|t)$ and $p(W|t)$ 
are obtained approximating the distributions as independent Gaussians and using the relations between age and the Gaussian parameters found in Section \ref{sec:method_UVW}:

\begin{equation}
\label{eq:prob_t_method2}
p(V|t) \cdot p(W|t) = \frac{1}{2\pi} \, \mathlarger{\mathlarger{\prod_{\mathsmaller{\mathsmaller{k = V, W}}}}}\left[\frac{1}{\sigma_k(t)} \, \exp{\left( - \frac{\left[k + k_{\sun}(t)\right]^2}{2\,\sigma_k^2(t)} \right)}\right]
\end{equation}


\subsection[Method eUW]{Method $\boldsymbol{eUW}$}

Method $eUW$ is essentially the same as Method $eVW$, but for this method, we are considering the $U$ velocity component instead of the $V$ component. In this case, the density maps needed are the ones in the $eU$ plane. The density maps were build from the same simulated data used for the ones described in Method $eVW$. The Bayesian equation that gives the age probability from the kinematical parameters is

\begin{equation}
p(t|e,U,W) = p(e|U,t) \, p(U|t) \, p(W|t) \, p(t) \,.
\end{equation}

As before, we obtain the age pdf calculating the probability $p(t|e,U,W)$ for different ages. In each case, $p(e|U,t)$ is obtained through the density maps and $p(U|t)$ and $p(W|t)$ are obtained approximating gaussians and using the relations found for the parameters in Section \ref{sec:method_UVW}:

\begin{equation}
\label{eq:prob_t_method3}
p(U|t) \cdot p(W|t) = \frac{1}{2\pi} \, \mathlarger{\mathlarger{\prod_{\mathsmaller{\mathsmaller{k = U, W}}}}}\left[\frac{1}{\sigma_k(t)} \, \exp{\left( - \frac{\left[k + k_{\sun}\right]^2}{2\,\sigma_k^2(t)} \right)}\right]
\end{equation}

\section{Discussion}
\label{sec:discussion}
We have applied the methods described in Section \ref{sec:methods} to derive the pdf for the age of all $9\,102$ stars from the sample defined in Section \ref{sec:sample}. Nine out of these stars could not have their ages determined by Methods $eVW$ and $eUW$ because their velocities were higher than the range we considered while building the density maps (i.e., $\vert U \vert \, \mathrm{or} \, \vert V \vert > 200 \kms$). As most of these stars also have high $W$ velocity, they should belong to other components of the Galaxy rather than the thin disk and may not affect our conclusions.

While it is best to use all information available in the posterior age pdf, it is often necessary to use a single age estimator in order to study the relation of the stellar parameters and age. As our first goal is to compare the results obtained from the kinematical method to those obtained from the isochronal one, we use the most likely age $t_{\mathrm{ML}}$ and expected age $t_{\mathrm{E}}$ obtained from the pdf to characterize single ages:

\begin{subequations}
\begin{align}
	t_{\mathrm{ML}} &= \argmax_t f(t) \,, \\
	t_{\mathrm{E}} &= \int_{-\infty}^{\infty} \! \! \! \! t \, f(t) \, \mathrm{d}t \,,
\end{align}
\end{subequations}
where $f(t)$ represents the posterior probability density function $p(t|U,V,W)$, $p(t|e,V,W)$ or $p(t|e,U,W)$, respectively for Methods $UVW$, $eVW$ and $eUW$.

\begin{figure}
\centering
\includegraphics[scale=1]{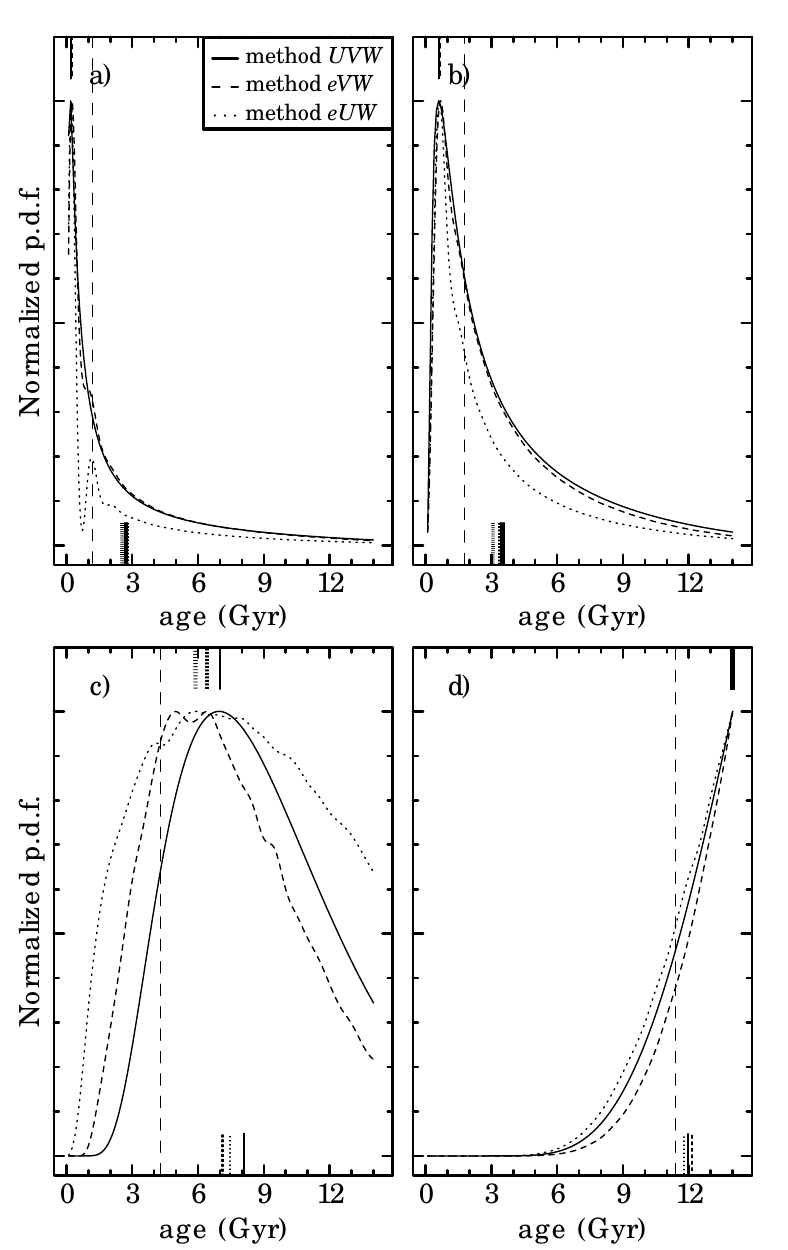}
\caption{Probability density function obtained by Method $UVW$ (solid line), Method $eVW$ (dashed line) and Method $eUW$ (dotted line), for four representative stars. Panel a) HD 1101, which has velocities $(U,V,W) = (-12, -23, -6) \kms$ and eccentricity $e = 0.07$, b) HD 1343, for which the velocities are $(U,V,W) = (-22, 16, 3) \kms$ and the eccentricity is $e = 0.09$, c) HD 852, that has $(U,V,W) = (-58, 39, -9) \kms$ and $e = 0.22$ and d) HD 12387, with velocities $(U,V,W) = (-3, -88, 60) \kms$ and eccentricity $e = 0.32$. The vertical marks at the top axis represents the most likely age for each method and those at bottom axis represents the expected ages.}
\label{fig:pdf_example}
\end{figure}

Figure \ref{fig:pdf_example} shows examples of pdfs obtained for four different stars, using the three methods previously described. Also plotted are the most likely age (upper axis lines) and the expected age obtained for each method (lower axis lines). These stars were selected as examples of the different situations one can find when calculating the pdf. In panel a), we can see that there is no defined lower limit for the stellar age, therefore, the most probable age for this star is defined as the lowest age considered in this work (0.1 Gyr). Panel b) represents the case for which both the expected age, and the most likely age can be properly defined. In the case of Panel c), even though a most likely age can be obtained, the expected age will be biased towards lower ages because of the truncation our prior $p(t)$ imposes on the pdf. Panel d) represents a case similar to Panel a), but in this case the limitation is caused by the fact that the most likely age would be greater than the age of the universe.

Since removing all stars whose age pdfs happens to be the cases a), c) or d) would seriously diminish our sample, we have chosen instead to work with a combination of the expected age ($t_{\mathrm{E}}$) and the most likely age ($t_{\mathrm{ML}}$). This way, the sample is less affected by the bias discussed above as in most cases at least one of the ages is well defined. We then define the kinematical age $t_{\mathrm{kin}}$ as a weighted average between $t_{\mathrm{ML}}$ and $t_{\mathrm{E}}$. We found that the expression giving the most similar results as isochronal ages is:
\begin{equation}
\label{eq:t_kin_definition}
t_{\mathrm{kin}} = \frac{3\,t_{\mathrm{ML}} + t_{\mathrm{E}}}{4} \mathrm{.}
\end{equation}

To tabulate the results, we have calculated from each stellar age pdf the expected age ($t^{(i)}_{\mathrm{E}}$), the most likely age ($t^{(i)}_{\mathrm{ML}}$) and the ages corresponding to the $2.5\%$, $16\%$, $50\%$, $84\%$ and $97.5\%$ percentiles, respectively $t^{(i)}_{2.5}$, $t^{(i)}_{16}$, $t^{(i)}_{50}$, $t^{(i)}_{84}$ and $t^{(i)}_{97.5}$, where $i = 1, 2, 3$ designates the method used to derive the age pdf, respectively methods $UVW$, $eVW$ and $eUW$. The results for the first 75 sample stars are shown in Table \ref{tab:results50}. The full data sample is only available in electronic format.

\subsection{Uncertainties for individual ages}
\label{sec:uncertainties}

Due to the statistical nature of this method, uncertainties arise from two different sources: (i) the observational uncertainties that affects the $UVW$ values and, consequently, the resulting pdf and (ii) uncertainties caused by our lack of knowledge of the exact value of the Solar peculiar velocity. We analyze how the latter affects the results by changing the Solar motion and, as the former depends on the quality of the data, we investigate how the results change when considering improved observational precision.

\subsubsection{Uncertainties due to statistical nature of the Method}

\begin{figure}
\centering
\includegraphics[scale=1]{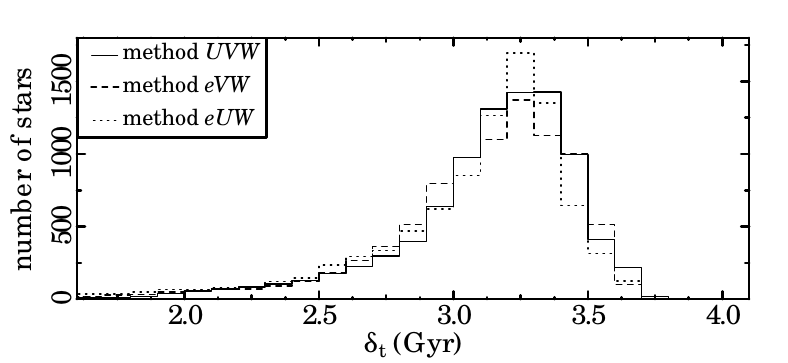}
\caption{Distribution of individual uncertainties for the ages obtained by Method $UVW$ (solid line), Method $eVW$ (dashed line) and Method $eUW$ (dotted line) using Equation \ref{eq:delta_t_definition}. All methods are very similar and the $\delta_t$ distributions peak at about 3.25 Gyr, with no uncertainty higher than 4 Gyr.}
\label{fig:dist_delta}
\end{figure}

To estimate the uncertainties, related to the pdf spread, for the ages obtained for single stars through Equation \ref{eq:t_kin_definition}, we have used an expression involving the ages of the percentiles described above. The expression for the uncertainty was chosen in a way that, for a Gaussian distribution, $\delta_t$ would correspond to 1 sigma, and is given by
\begin{equation}
\label{eq:delta_t_definition}
\delta_t = \frac{1}{4} \left[ (t_{84} - t_{16}) + \frac{t_{97.5} - t_{2.5}}{2} \right] \, \mathrm{.}
\end{equation} 
This definition is a measure of the spread that takes into account both the dispersion closer to the median age (from $t_{84}$ and $t_{16}$) and the dispersion at the tails of the distribution (from $t_{2.5}$ and $t_{97.5}$). It is designed to allow a direct comparison between the spread of the calculated kinematical method and original isochronal pdf obtained by \citet{Casagrande+2011}, whose data contains the data for these percentiles. It's important to keep in mind that, although a valid measure of the spread, it's arbitrary and not always corresponds to the Gaussian $1\sigma$, as the pdf can be very different from a normal distribution.

The distribution of individual uncertainties for each of the three methods is represented in Figure \ref{fig:dist_delta}. It can be seen that, concerning individual uncertainties, the methods presented here are very similar to each other. Figure \ref{fig:dist_delta} also shows that, for individual stars, uncertainties are very high. The mean uncertainty for Methods $UVW$ and $eVW$ is 3.1 Gyr, and for Method $eUW$, 3.0 Gyr. For comparison, the mean uncertainty for isochronal ages, defined in a similar way, would be about 0.7 Gyr for this sub-sample of the GCS. This means the kinematical ages are useful for individual stars only when the isochronal method is certain to result in extremely high uncertainties, as is the case for M dwarfs, or as an independent age indicator that can complement other methods.

\subsubsection{Impact of observational uncertainties}

In order to understand how the observational uncertainties that leads to the calculation of the $UVW$ velocities affect the obtained pdf, we have performed a set of Monte Carlo simulations and analyzed the effect of this in the calculated $t_\mathrm{ML}$ and $t_\mathrm{E}$.

For each star, the parallax, proper motions and radial velocities were re-sampled 1000 times, considering Gaussian individual errors, and the $UVW$ velocities were recalculated in each case.

\begin{figure}
\centering
\includegraphics[scale=0.52]{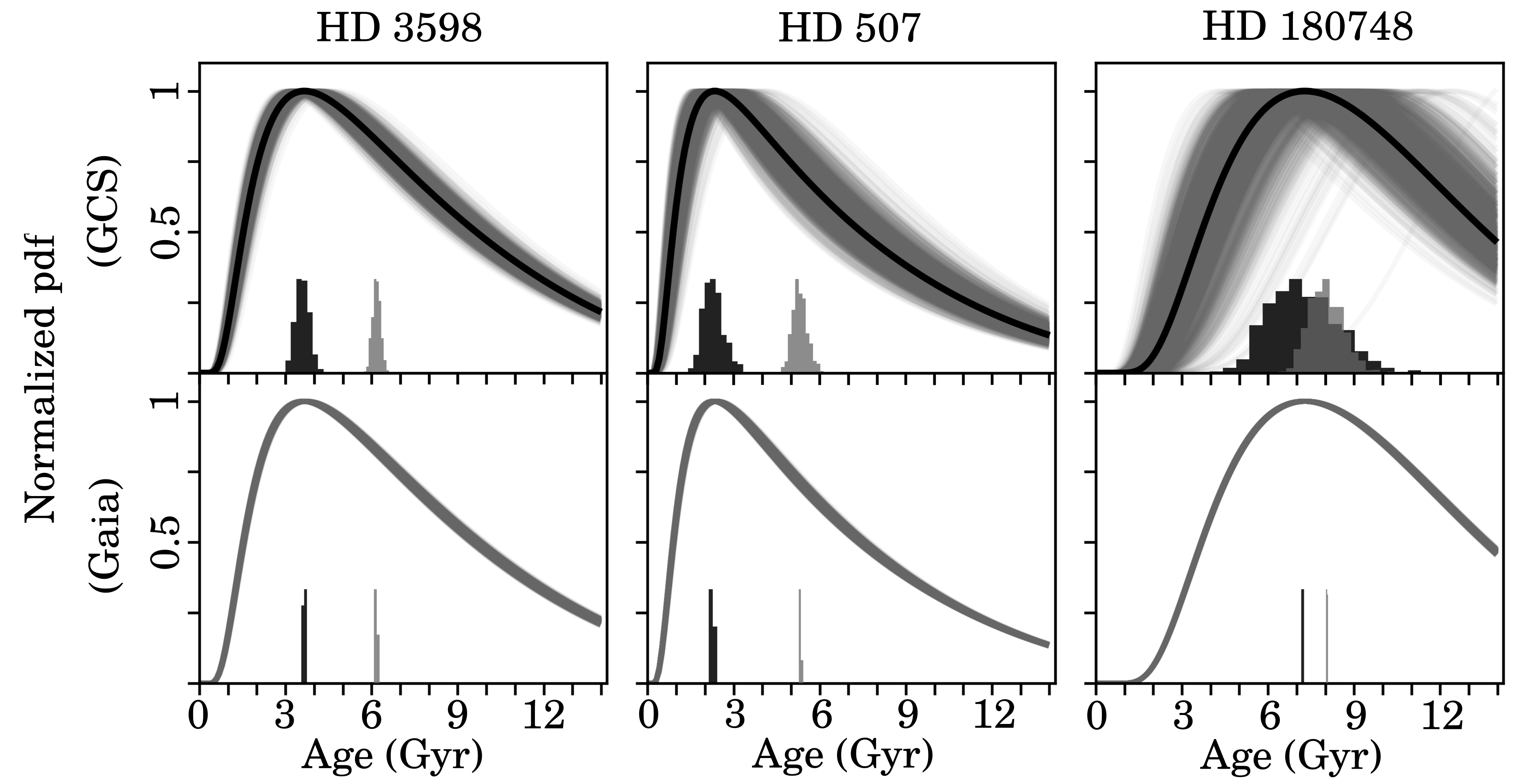}
\caption{Effects of the observational uncertainties on the pdf considering the GCS uncertainties (top) and Gaia's predicted uncertainties (bottom) for the stars HD 3598 (left), HD 507 (middle) and HD 180748 (right). The black solid line represents the original pdf, while the grey lines are the results of 1000 Monte Carlo simulations. The dark grey histogram represents the distribution of most-likely ages and the light grey histogram, the distribution of expected ages.}
\label{fig:pdf_monte_carlo}
\end{figure}

The top panels of Figure \ref{fig:pdf_monte_carlo} shows the pdfs obtained for each of the 1000 Monte Carlo simulations (grey lines) and the original pdf (black line) for three representative star: HD 3598 (left) which has uncertainties considered small for the GCS survey ($\sigma_\pi = 0.53$ mas, $\sigma_\mu = 1.0$ mas/year and $\sigma_{r_v} = 0.3 \, \kms$); HD 578 (middle) which has uncertainties representative of the most stars in the survey ($\sigma_\pi = 0.66$ mas, $\sigma_\mu = 2.0$ mas/year and $\sigma_{r_v} = 0.4 \, \kms$); and HD 180748 (right) which has large uncertainties compared to other stars in the survey ($\sigma_\pi = 1.8$ mas, $\sigma_\mu = 4.0$ mas/year and $\sigma_{r_v} = 3 \, \kms$). Also represented in each plot, is the distribution of the obtained $t_\mathrm{ML}$ (dark histograms) and $t_\mathrm{E}$ (light grey histograms). The dispersions of the histograms are: $\sigma_{t_\mathrm{ML}} = 0.26$ Gyr and $\sigma_{t_\mathrm{E}} = 0.15$ for HD 3598; $\sigma_{t_\mathrm{ML}} = 0.39$ Gyr and $\sigma_{t_\mathrm{E}} = 0.27$ for HD 578; and $\sigma_{t_\mathrm{ML}} = 1.25$ Gyr and $\sigma_{t_\mathrm{E}} = 0.61$ Gyr for HD 180748.

It's clear that, the higher the observational uncertainties are, the greater is the uncertainties in the calculated age point estimator. For the whole sample, considering the GCS uncertainties, we obtain on average $\sigma_{t_\mathrm{ML}} = 0.47$  and $\sigma_{t_\mathrm{E}} = 0.28$. Therefore, considering the observational uncertainties of the GCS, the expected age is less affected by uncertainties than the most likely age.

Also, to verify how much the upcoming Gaia data will improve the results, we perform another set of Monte Carlo simulations considering the uncertainties that Gaia aims to achieve by the end of the mission \citep{Gaia2005}: $\sigma_\pi = 25 \cdot 10^{-3}$ mas, $\sigma_\mu = 13.15 \cdot 10^{-3}$ mas/year and $\sigma_{r_v} = 1 \, \kms$. The lower panels of Figure \ref{fig:pdf_monte_carlo} shows the obtained pdfs for the stars HD 3598 (left), HD 578 (middle) HD 180748 (right). As can be seen, the quality of the expected Gaia data practically eliminates the intrinsic uncertainty of the method due to observational uncertainties. On average, for the whole sample, we obtain $\sigma_{t_\mathrm{ML}} = 0.09$ Gyr and $\sigma_{t_\mathrm{E}} = 0.07$ Gyr.

\subsubsection{Effect of changing the Solar peculiar velocity}
\label{sec:uncertainties-Solar-motion}

\begin{figure}
\centering
\includegraphics[scale=0.52]{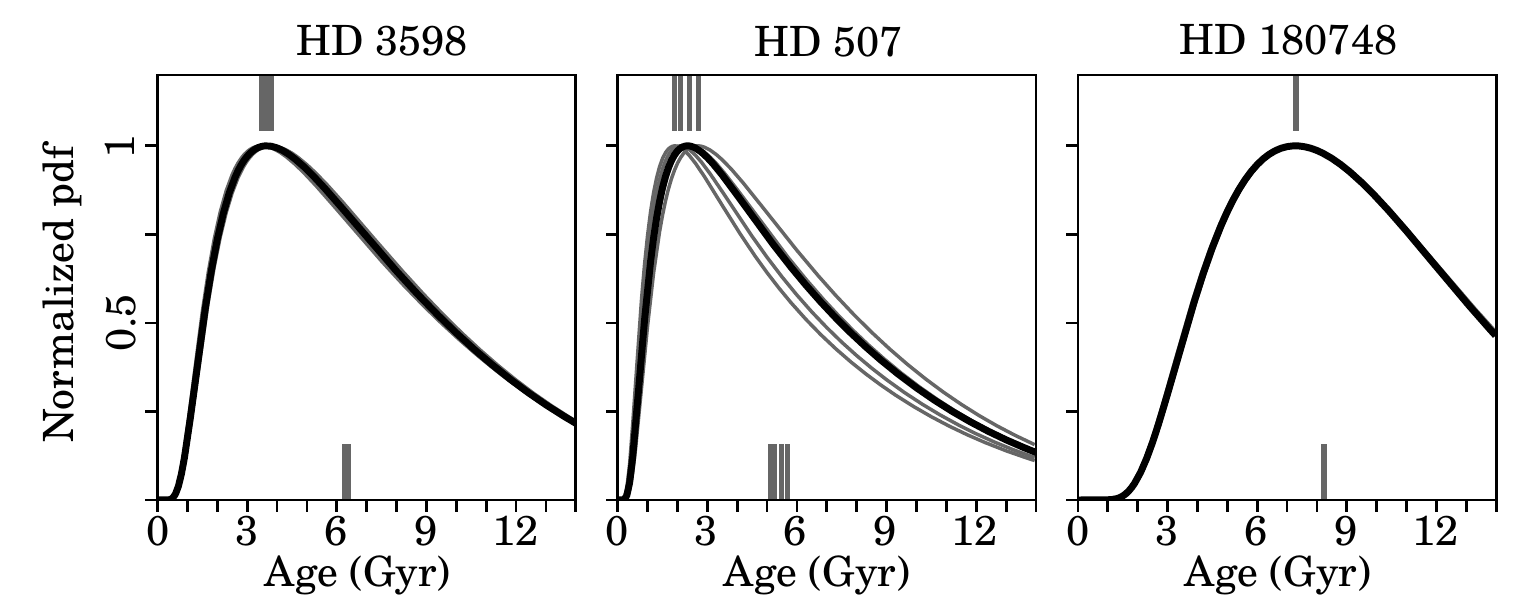} 
\caption{Effects of changing the Solar peculiar velocity on the calculated pdfs of the stars HD 3598 (left), HD 507 (middle) and HD 180748 (rigth). The black line represents the pdf obtained using the Solar motion derived in this work, while the grey lines represents the pdfs obtained using the Solar velocities displayed in Table \ref{tab:sun_uvw}.}
\label{fig:pdf_solar_velocity}
\end{figure}

We have also investigated the impact of changing the Solar peculiar velocity in the obtained pdfs and consequently in the estimated values of $t_\mathrm{ML}$ and $t_\mathrm{E}$. Figure \ref{fig:pdf_solar_velocity} shows the pdf obtained for the solar velocity obtained in this work (black line) and the pdfs obtained using the other values of solar velocity shown in Table \ref{tab:sun_uvw} (grey lines). We see that, considering the GCS uncertainties, the results are less affected by changing the Solar peculiar velocity than by the observational uncertainties (which will no longer be the case for the data quality expected by Gaia).

Compared to the Solar velocity of this work, the standard deviation of the age differences of the whole sample are: $\sigma_{t_\mathrm{ML}} = 0.26$ Gyr and $\sigma_{t_\mathrm{E}} = 0.18$ Gyr for the Solar peculiar velocity of F\&A14; $\sigma_{t_\mathrm{ML}} = 0.25$ Gyr and $\sigma_{t_\mathrm{E}} = 0.18$ Gyr for B\&B14; $\sigma_{t_\mathrm{ML}} = 0.21$ Gyr and $\sigma_{t_\mathrm{E}} = 0.14 $ for Co{\c s}+11, $\sigma_{t_\mathrm{ML}} = 0.08$ Gyr and $\sigma_{t_\mathrm{E}} = 0.06 $ for Sch+10 and $\sigma_{t_\mathrm{ML}} = 0.52$ Gyr and $\sigma_{t_\mathrm{E}} = 0.31 $ for Kov+09.

\subsection{Comparison between different methods}

Our method uses the isochronal ages derived from the Padova grid by \citet{Casagrande+2011} to fit the relations between the kinematical parameters and age, which are then used in Equations \ref{eq:v1v2v3}, \ref{eq:prob_t_method1}, \ref{eq:prob_t_method2} and \ref{eq:prob_t_method3}. To obtain a more independent comparison, we compare our derived kinematical ages against those derived by \citet{Casagrande+2011} using the BaSTI grid \footnote{While using the BaSTI ages for comparison is certainly better than using the Padova ages (which were used to fit the kinematical equations) they can't be seeing as fully independent age estimators. One of the criteria used to select the stars with well determined ages was that Padova and BaSTI ages had to agree within 1 Gyr.}. As these isochronal ages are also derived from a probability density function, we defined individual isochronal ages in the same way as before:
\begin{equation}
\label{eq:t_iso_definition}
t_{\mathrm{iso}} = \frac{3\,t^{(\mathrm{iso})}_{\mathrm{ML}} + t^{(\mathrm{iso})}_{\mathrm{E}}}{4} \, \mathrm{,}
\end{equation}
where $t^{(\mathrm{iso})}_{\mathrm{ML}}$ is the most likely age and $t^{(\mathrm{iso})}_{\mathrm{E}}$ is the expected age derived from the age pdf.

\begin{figure}
\centering
\includegraphics[scale=1]{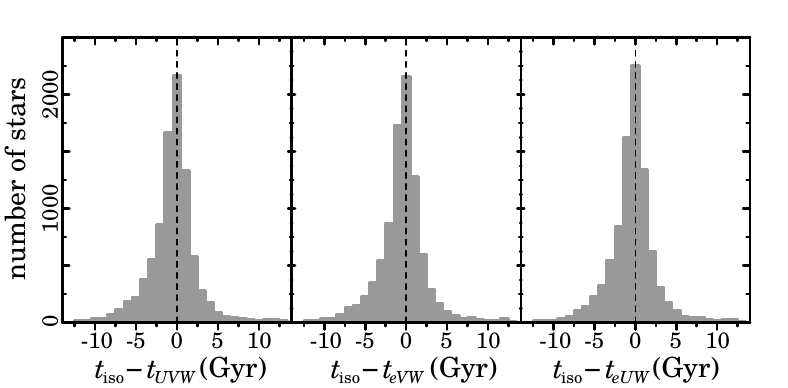}
\caption{Distribution of difference between isochronal age and kinematical age, obtained by Method $UVW$ (left), Method $eVW$ (middle) and Method $eUW$ (right). The distributions are similar for all three methods and peak at zero, showing that the ages display a tendency of good agreement. The spreads can be explained by the individual uncertainties of both isochronal and kinematical methods.}
\label{fig:Delta_iso_kin}
\end{figure}

We plotted in Figure \ref{fig:Delta_iso_kin} the distributions of the differences between the isochronal and the kinematical ages derived by the three Methods. The behaviour of all distributions is similar, the peak is close to zero, meaning that the most common case is an agreement between isochronal and kinematical ages; the median of the distributions is also close to zero, being $-$0.33 for the case of Method $UVW$ (left), $-$0.34 for Method $eVW$ (middle) and $-$0.24 for Method $eUW$ (right). The spread in the distribution is explained by the high uncertainties of the kinematical methods, coupled with the also considerable uncertainties of the isochronal method. 

Also noticeable is the long tail towards negative values that appears in all distributions, meaning that, in some cases, kinematical ages might be overestimated. This can be explained by the large spread in the pdf, which pushes the expected age towards the center of the age interval. As there are more young than old stars in the sample, there are more stars having their expected age overestimated than underestimated. The reason we chose to give less weight to the expected age in our kinematical age definition, Eq. \ref{eq:t_kin_definition}, was to reduce this effect.

\begin{figure}
\centering
\includegraphics[scale=1]{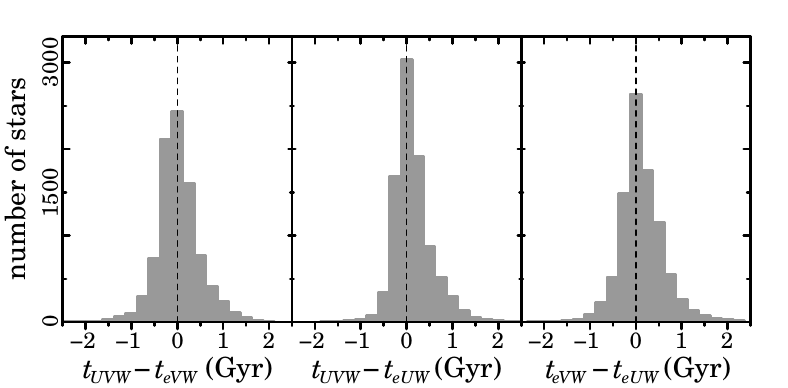}
\caption{Distributions of differences between ages obtained by the different kinematical methods. In all cases the peak of the distribution is close to zero, and the differences rarely exceed 1 Gyr, showing good agreement between the methods.}
\label{fig:Delta_kin_kin}
\end{figure}

We have also compared the ages obtained by the different methods present here, as can be seen in Figure \ref{fig:Delta_kin_kin}. There is very good agreement between all Methods as the differences rarely exceed 1 Gyr. The peak of the differences distributions is also very close to zero, the mean being 0.02 in the case of Methods $UVW$ and $eVW$, 0.14 for Methods $UVW$ and $eUW$ and 0.12 for Methods $eVW$ and $eUW$.

\subsection{Age distribution}

\begin{figure}
\centering
\includegraphics[scale=1]{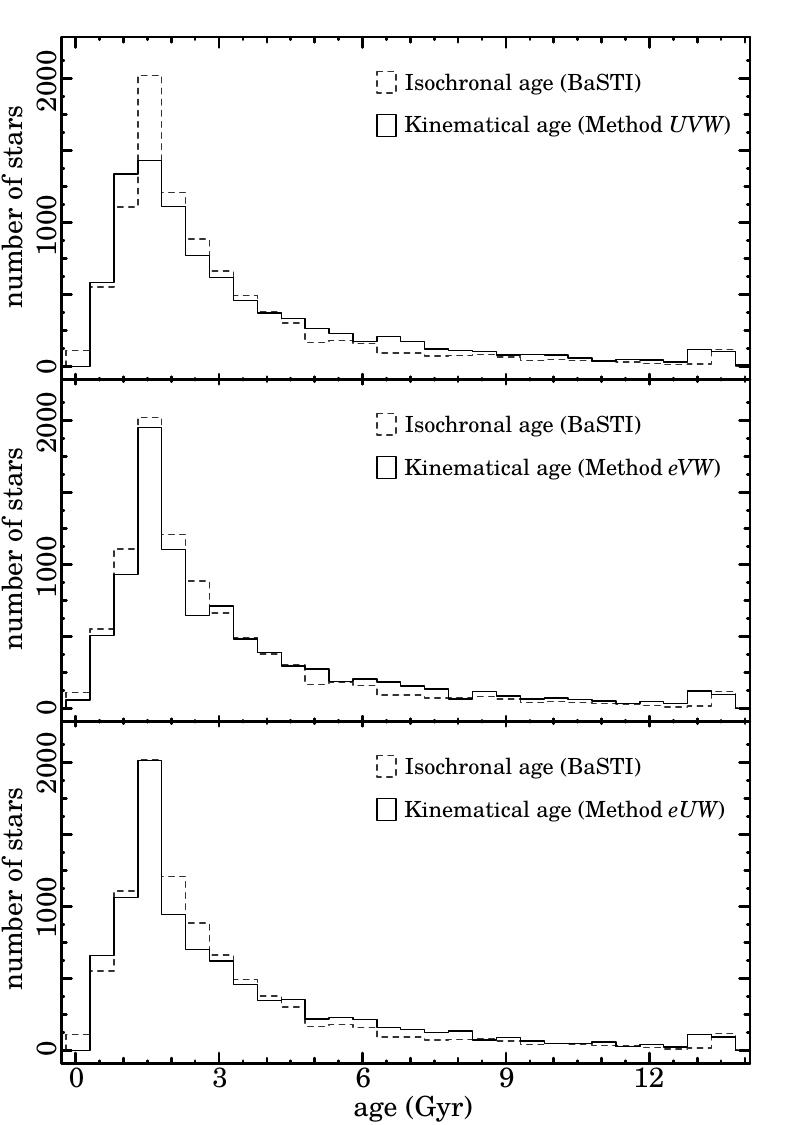}
\caption{Distribution of ages obtained by Method $UVW$ (top), Method $eVW$ (middle) and Method $eUW$ (bottom). For comparison, the distribution of isochronal ages is plotted in all panels (dashed line). As can be seen, the ages distributions obtained by the kinematical methods are very similar to the ones obtained by the isochronal case. The agreement is even better for the cases that include eccentricity.}
\label{fig:age_dist}
\end{figure}

As the individual uncertainties $\delta_t$ are too high we only recommend this analysis for individual stars as an independent age indicator or when other precise methods are not available. Even though, the kinematical age method may be useful when the properties of large samples are considered. One of its applications is to derive the age distribution of stars in a given sample. Figure \ref{fig:age_dist} shows the distributions of stellar ages obtained by Method $UVW$ (top), Method $eVW$ (middle) and Method $eUW$ (bottom). Also, for comparison, we plotted the isochronal age distribution in all panels. In all cases, especially for Methods $eVW$ and $eUW$, the derived age distribution agrees very well to the isochronal case.

The age distribution in a sample is an important parameter, closely related to the star formation history in the Galaxy. The best stars to observe for this task are the ones that evolves slowly enough so that those born in the early phases of our Galaxy formation are still around. While slow evolution is what makes these stars live long enough to trace the Galactic history, it also makes it harder to obtain the stellar ages through their internal properties since they slowly change with time. This slow internal evolution does not affect the evolution of the parameters used by the kinematical method (spatial velocities and eccentricity), which varies with time in the same manner for star of different masses. Therefore, the kinematical method has a role in helping the understanding of important aspects of the evolution of our Galaxy.

In this work, we simply compare the kinematical age distribution with the isochronal age distribution. This distribution cannot be directly interpreted as a star formation history because the selection of stars for the Geneva-Copenhagen Survey \citep{Nordstrom+2004} was based in photometric cuts, which favours certain regions of the $\Teff-M_{\mathrm{bol}}-[\mathrm{Me}/\mathrm{H}]$ space, directly biasing the age distribution. To translate the age distribution to star formation history, a correction for this bias must be applied, which is beyond the scope of this work.
 
\subsection{Age-metallicity relation}

The chemical evolution of the Galaxy remains largely unknown. Studies directly relating chemical abundances of different elements to specific epochs are still restricted to very few stars \citep{Spina+2016}. Therefore, to understand this evolution with the available data, we have to rely on models and observational constraints.

The first models considered a simple closed box scenario \citep{Talbot+Arnett1971} in which stars expel their material enriching the interstellar medium, causing newborn stars to be more metal rich than previous generations. The metallicity distribution predicted for present-day stars from this model does not agree well with observations \citep[e.g. the G-dwarf problem][]{Schmidt1963, Pagel+Patchett1975, Wyse+Gilmore1995, Rocha-Pinto+Maciel1996, Haywood2001, Nordstrom+2004}, indicating that more complex processes are involved. When inflows and outflows of gas \citep{Larson1972, Hartwick1976, Schindler+Diaferio2008}, and radial stellar migration \citep{Wielen+1996, Sellwood+Binney2002, Minchev+2013} are considered, the agreement between model predictions and observations is remarkably enhanced.

Considering the complexity of the models, more observational constraints are required. One property that can elucidate aspects of the Galactic chemical evolution is how the overall metallicity of stars changes with time, the so called age-metallicity relation (AMR). The most critical part of obtaining the AMR is determining stellar ages for low mass stars, especially the ones that can live enough to tell us about the early history of the Galaxy. The feature of the AMR that models must be able to explain is the low \citep[or none at all, see][]{Edvardsson+1993, Nordstrom+2004} increase of metallicity with time and the increase of the metallicity dispersion with age \citep{Casagrande+2011}.

\begin{figure}
\centering
\includegraphics[scale=1]{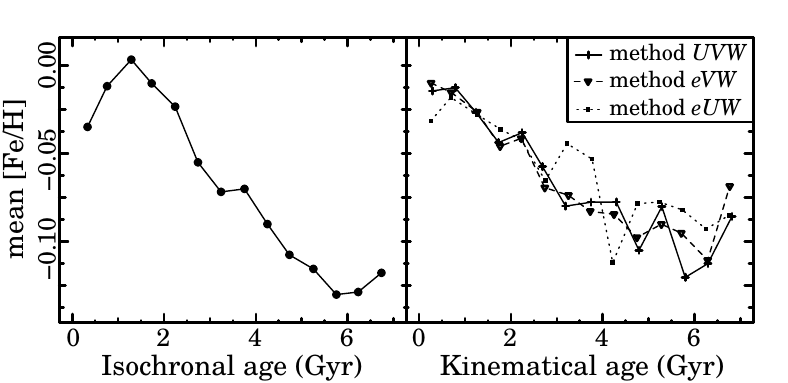}
\caption{Relation between the mean metallicity and the ages, lower than 7 Gyr, obtained through isochronal method (left) and kinematical methods (right). The behaviour of the relation is similar in both cases, suggesting that the kinematical method is useful for finding relations between chemical features and age.}
\label{fig:AMR}
\end{figure}

To check whether the kinematical method can be useful to derive the age-metallicity relation, we divided our stellar sample by bins of age 0.5 Gyr wide and calculated the mean $[\mathrm{Fe}/\mathrm{H}]$ for each group. The same was done for the isochronal ages for comparison. As groups older than 7 Gyr would have very few stars, we have restricted our analysis for ages lower than this. The results are plotted in Figure \ref{fig:AMR}. This raw analysis of the data shows a decay of metallicity with age both for the isochronal and the kinematical cases. The relation is slightly flatter for the kinematical ages. This is probably caused by the higher uncertainties for individual ages, which causes a mixture between stars that should, in reality, belong to other age bins.
 
As there is also a bias caused by selection criteria, we do not interpret the result as a true age-metallicity relation in the Solar Neighbourhood. In this work, we are just interested in showing that the results obtained from the kinematical ages are similar to those obtained using isochronal ones, therefore concluding that the kinematical method might be used for this task for samples of stars for which ages cannot be determined by isochrones.

\subsection{The age of Trappist-1}
\label{sec:trappist1}

Trappist-1 is the system that hosts the largest number of known earth-size planets in its habitable zone \citep{Gillon+2016}. Recent results show that the star harbors 5 planets of sizes similar to that of the Earth (planets b, c, e, f and g) and 2 planets slightly smaller ($\approx0.75$ $R_{\mathrm{\oplus}}$, planets d and h; \citealp{Gillon+2017}). The planets e, f and g are in the habitable zone and, therefore, could harbour water oceans.

After its discovery in 2016 and confirmation of 3 earth-size planets in the habitable zone in 2017, this star received a lot of attention and has been the central investigation of several studies consisting of further analysis of light curves \citep[e.g.][]{Wang+2017}, dynamical investigations \citep[e.g.][]{Tamayo+2017, Quarles+2017}, and habitability plausibility \citep[e.g.][]{OMalley+Kaltenegger2017, Wolf2017}. Despite of this, one very important information is still missing: the stellar age.

The reason for the lack of age information is the fact that Trappist-1 has a mass of only 8\% the mass of the Sun. It causes the stellar evolution to be too slow, making the atmospheric parameters practically unchanged since the stellar birth. Because of this, most of the traditional age dating methods can not be applied. Since the kinematical method is expected to be independent from the internal evolution of the star and depends only on its orbital evolution, we can apply the method to obtain the age p.d.f. for the star and estimate its expected value and uncertainty.

\begin{figure}
\centering
\includegraphics[scale=0.75]{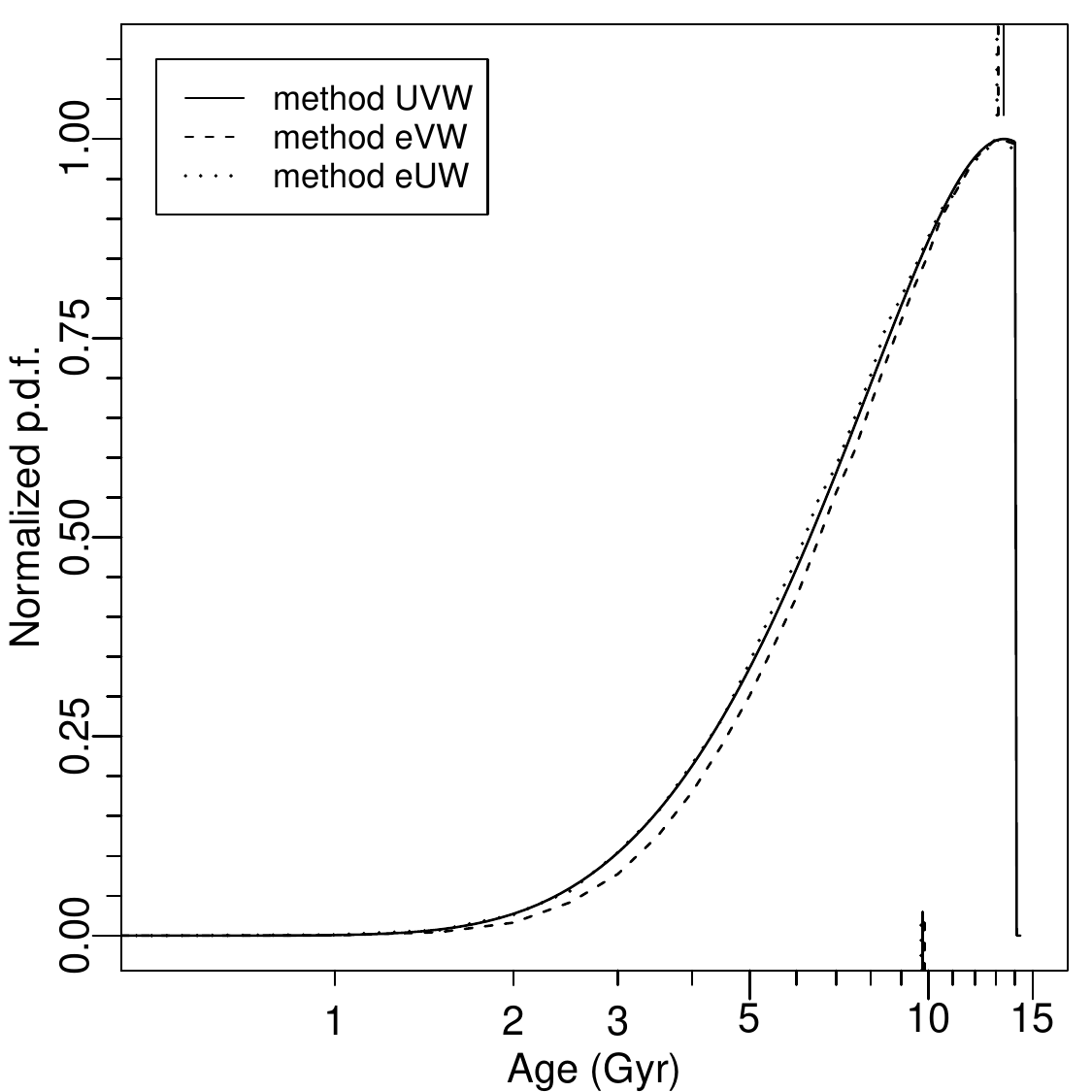}
\caption{Age pdfs obtained for the Trappist-1 star using Method UVW (solid line), Method eVW (dashed line) and Method eUW (dotted line). The vertical lines in the top axis represent the most-likely age and the lines in the bottom axis, the expected age.} 
\label{fig:Trappist_age}
\end{figure}

From the stellar coordinates, proper motions and parallax \citep{Costa+2006} and radial velocity \citep{Burgasser+2015}, we derived its heliocentric peculiar velocities using the method described by \citet{Johnson+Soderblom1987}. The obtained values were $U = -43.24 \kms$, $V = -66.25 \kms$ and $W = 13.87 \kms$. The eccentricity was calculated using Equation \ref{eq:ecc} and found to be 0.27. We used these velocities and eccentricity to apply Methods UVW, eVW and eUW to derive the age pdfs represented in Figure \ref{fig:Trappist_age}.

Figure \ref{fig:Trappist_age} shows that the most likely age (vertical marks at the top axis) and the expected age (vertical marks at the lower axis) for the three different methods. The most-likely ages for methods $UVW$, $eVW$ and $eUW$ are respectivelly 13.4, 13.13 and 13.06 Gyr and the expected ages are respectivelly 9.78, 9.86 and 9.73, showing good agreement between the three kinematical methods. We then apply Equation \ref{eq:t_kin_definition} to calculate the defined kinematical ages, and obtain 12.50 for Method $UVW$, 12.31 for Method $eVW$ and 12.23 for Method $eUW$.

In order to estimate upper and lower limits, we also calculate the 16\% and 84\% percentile ages, and obtained $t_\mathrm{kin}^{(UVW)} = 12.50^{+0.29}_{-6.23}$, $t_\mathrm{kin}^{(eVW)} = 12.31^{+0.53}_{-6.05}$, $t_\mathrm{kin}^{(eUW)} = 12.23^{+0.56}_{-6.22}$. In this case the kinematical Method only provides a good upper-limit for the age and has a large tail towards lower ages, as can also be seen in Figure \ref{fig:Trappist_age}. The results allows us to conclude that the age of the system most-likelly lies between $\approx$6--12.5 Gyrs. Considering the observational uncertainties, the results are in good agreement with the age derived by \citet{Burgasser+Mamajek2017}: $7.6 \pm 2.2$ Gyr. In this case, the authors have considered several age constrains: the stellar cmd, average density, lithium absorption, surface gravity, rotation, magnetic activity, as well as kinematics.

We believe the derived age further increases the astrobiological interest for the system: it not only contains 7 planets with sizes similar to the earth (3 of which are in the habitable zone) but is also old enough for life as we know it to develop and evolve. Nevertheless an important remark must be made: as the system is extremely compact (the highest period being of 18.77 days; \citealp{Luger+2017}) the planets are all expected to be tidally locked \citep{Gillon+2017}, causing the temperatures to vary significantly between the face towards the star and the opposite one, which would constitute another barrier for the formation of life.

\section{Conclusions}
\label{sec:conclusions}
We showed how a probability density function (pdf) for the age of a star may be obtained from its spatial velocity components $U$, $V$ and $W$ and also from other kinematical parameters like the eccentricity. We characterize individual ages from the pdf using the most likely and the expected ages. This individual age estimation has uncertainties of about 3 Gyr, which, although higher than classical methods, may be the best estimate available for most very low mass stars.

The methods are based on the growth of velocity dispersion with age. We adopted a relation between age and velocity dispersion as $\sigma_i(t) = b_i \, t^{a_i}$. The parameters $b_i$ and $a_i$ were found using a sub-sample of the Geneva-Copenhagen Survey, which contains isochronal ages derived by \citep{Casagrande+2011}. We show that other parameters that define the velocity distribution are also a function of stellar age: the $V$ component of the Solar motion ($V'_{\sun}$) and the vertex deviation ($\lv$). From this sample, we obtained the solar motion to be $(U,V,W)_{\sun} = (9.8\pm0.3, 12.5\pm0.9, 7.2\pm0.2)_{\sun} \kms$. Using the example of eccentricity, we show how other kinematical parameters may be included in the analysis, provided the density of stars in the kinematical parameter-velocity space is known for all ages.

We applied the Kinematical Method for the stars of the Geneva-Copenhagen survey and compared the results with those obtained by isochronal estimation. We show that differences between the obtained ages peak at zero and, despite a slightly overestimation of kinematical ages, there is no significant bias. This means that, even though uncertainties are large, the kinematical method is useful to derive statistical parameters for groups of stars. A comparison between the kinematical methods based purely on spatial velocity and the ones that include eccentricity shows that there is a very good agreement in the ages derived and the differences rarely exceed 1 Gyr.

We investigated how the results are affected by observational uncertainties and by changing the Solar peculiar velocity. We conclude that, for the GCS uncertainties, the results are more affected by the observational uncertainties than by changes in the Solar peculiar velocity, but it will no longer be the case for the data quality that Gaia aims to achieve.

The distribution of stellar ages obtained through the kinematical methods agrees very well with the one obtained through isochrones, principally for the methods that include the eccentricity. We have not attempted to derive the star formation history from this data, because it would require the consideration of a detailed analysis of the age bias imposed by the sample selected criteria, which is left for future work. However, the agreement between kinematical and isochronal distribution leads us to conclude that the kinematical method may be as useful as the isochronal for this task, with the advantage of being applicable to the numerous very low mass stars.

The relation between metallicity and age, which is an important constraint for chemical evolution models, has also been investigated using the kinematical method. The behaviour of the relation shown for stars younger than 7 Gyr, is similar to the one obtained through isochrones, suggesting the kinematical method can also be used to derive this relation. An interpretation of this relation as a true age-metallicity relation also needs a detailed analysis of bias imprinted by selection criteria.

We showed that the method can also be applied for M stars, as it does not depend on the stellar mass. As an example, we derived the age for the Trappist-1 system, which is the known system that hosts the largest number of earth-size planets (seven, three of which are in the habitable zone). Our analysis of the obtained pdfs indicated an age between $\approx$6--12.5 Gyr.

We intent to explore this method to obtain kinematical ages pdfs for stars in large surveys, such as the Radial Velocity Experiment (RAVE) \citep{Steinmetz+2006} and Gaia \citep{Gaia+2016}, in the cases the other methods cannot be applied or would be unreliable. These pdfs can also be used as an independent age indicator to corroborate the results obtained by other means.

\textit{Acknowledgements}
For providing support with a PhD grant, Almeida-Fernandes F. wants to thank CAPES --- The Brazilian Federal Agency for Support and Evaluation of Graduate Education within the Ministry of Education of Brazil, and for the support, when the author was an undergrad, Almeida-Fernandes F. wants to thank CNPq --- The Brazilian National Council for Scientific and Technological Development.



\bibliographystyle{mnras}
\bibliography{bibliography} 




\appendix

\section{Results for the first 75 stars}

\begin{table*}
	\scriptsize
	\centering
	\caption{Kinematical ages obtained through methods $UVW$ (1), $eVW$ (2) and $eUW$ (3) for the first 75 sample stars.}
	\label{tab:results50}
	\begin{tabular}{p{1cm}>{\raggedleft\arraybackslash}p{0.35cm}>{\raggedleft\arraybackslash}p{0.35cm}>{\raggedleft\arraybackslash}p{0.35cm}>{\raggedleft\arraybackslash}p{0.3cm}>{\raggedright\arraybackslash}p{0.22cm}>{\raggedright\arraybackslash}p{0.22cm}>{\raggedright\arraybackslash}p{0.22cm}>{\raggedright\arraybackslash}p{0.22cm}>{\raggedright\arraybackslash}p{0.22cm}>{\raggedright\arraybackslash}p{0.22cm}>{\raggedright\arraybackslash}p{0.22cm}>{\raggedright\arraybackslash}p{0.22cm}>{\raggedright\arraybackslash}p{0.22cm}>{\raggedright\arraybackslash}p{0.22cm}>{\raggedright\arraybackslash}p{0.22cm}>{\raggedright\arraybackslash}p{0.22cm}>{\raggedright\arraybackslash}p{0.22cm}>{\raggedright\arraybackslash}p{0.22cm}>{\raggedright\arraybackslash}p{0.22cm}>{\raggedright\arraybackslash}p{0.22cm}>{\raggedright\arraybackslash}p{0.22cm}>{\raggedright\arraybackslash}p{0.22cm}>{\raggedright\arraybackslash}p{0.22cm}>{\raggedright\arraybackslash}p{0.22cm}>{\raggedright\arraybackslash}p{0.22cm}}
\hline \\[-0.1cm]
Name & $U$ & $V$ & $W$ & $e$ & $t^{(1)}_{\mathrm{E}}$ & $t^{(1)}_{\mathrm{MP}}$ & $t^{(1)}_{2.5}$ & $t^{(1)}_{16}$ & $t^{(1)}_{50}$ & $t^{(1)}_{84}$ & $t^{(1)}_{97.5}$ & $t^{(2)}_{\mathrm{E}}$ & $t^{(2)}_{\mathrm{MP}}$ & $t^{(2)}_{2.5}$ & $t^{(2)}_{16}$ & $t^{(2)}_{50}$ & $t^{(2)}_{84}$ & $t^{(2)}_{97.5}$ & $t^{(3)}_{\mathrm{E}}$ & $t^{(3)}_{\mathrm{MP}}$ & $t^{(3)}_{2.5}$ & $t^{(3)}_{16}$ & $t^{(3)}_{50}$ & $t^{(3)}_{84}$ & $t^{(3)}_{97.5}$\\[+0.15cm]
\hline \\
HD 23        & 40 & -22 & -16 & 0.16 & 6.5 & 3.5 & 1.4 & 2.8 & 6.0 & 10.4 & 13.3 & 6.2 & 2.9 & 1.1 & 2.5 & 5.7 & 10.2 & 13.2 & 5.7 & 2.6 & 1.1 & 2.3 & 5.0 & 9.5 & 13.0\\
HD 105       & -10 & -21 & -1 & 0.06 & 3.1 & 0.3 & 0.1 & 0.4 & 1.8 & 6.4 & 11.9 & 3.5 & 0.6 & 0.3 & 0.7 & 2.3 & 6.9 & 12.0 & 3.4 & 0.6 & 0.2 & 0.6 & 2.2 & 6.7 & 12.0\\
HD 153       & -22 & -46 & 3.0 & 0.17 & 7.1 & 4.5 & 1.3 & 3.1 & 6.9 & 11.3 & 13.5 & 6.9 & 3.6 & 1.4 & 3 & 6.5 & 11.0 & 13.5 & 7.0 & 5.0 & 1.3 & 3.1 & 6.7 & 11.1 & 13.5\\
HD 156       & -9 & -1 & -15 & 0.02 & 3.3 & 0.5 & 0.2 & 0.6 & 2.1 & 6.4 & 11.7 & 3.0 & 0.6 & 0.3 & 0.7 & 1.9 & 5.7 & 11.1 & 3.1 & 0.6 & 0.3 & 0.6 & 1.9 & 5.9 & 11.3\\
HD 189       & 37 & -31 & -7 & 0.17 & 6.9 & 4.2 & 1.6 & 3.2 & 6.5 & 10.9 & 13.4 & 6.3 & 2.6 & 0.8 & 2.3 & 5.8 & 10.5 & 13.3 & 6.4 & 3.4 & 1.3 & 2.8 & 5.9 & 10.3 & 13.3\\
HD 200       & -4 & -50 & -23 & 0.18 & 8.1 & 7.6 & 2.2 & 4.3 & 8.1 & 11.9 & 13.6 & 7.9 & 7.5 & 2.1 & 4.2 & 7.9 & 11.8 & 13.6 & 7.8 & 6.6 & 2.1 & 4.1 & 7.7 & 11.7 & 13.6\\
HD 203       & -11 & -15 & -10 & 0.04 & 1.8 & 0.1 & 0.0 & 0.1 & 0.5 & 3.5 & 10.2 & 0.7 & 0.2 & 0.0 & 0.1 & 0.2 & 0.8 & 5.5 & 1.0 & 0.2 & 0.0 & 0.1 & 0.3 & 1.3 & 8.2\\
HD 268       & -28 & -50 & -18 & 0.19 & 7.7 & 6.6 & 1.6 & 3.7 & 7.6 & 11.7 & 13.6 & 7.4 & 5.4 & 1.7 & 3.6 & 7.3 & 11.5 & 13.6 & 7.5 & 5.3 & 1.6 & 3.6 & 7.4 & 11.5 & 13.6\\
HD 276       & -15 & -29 & -18 & 0.1 & 4.9 & 1.1 & 0.5 & 1.3 & 3.9 & 9.0 & 12.9 & 4.5 & 1.0 & 0.5 & 1.2 & 3.5 & 8.3 & 12.6 & 5.0 & 1.3 & 0.6 & 1.5 & 4.1 & 9.0 & 12.9\\
HD 285       & -50 & -10 & -15 & 0.12 & 5.2 & 1.9 & 0.8 & 1.8 & 4.4 & 9.1 & 12.9 & 5.1 & 1.6 & 0.7 & 1.7 & 4.2 & 8.9 & 12.8 & 5.0 & 1.7 & 0.8 & 1.7 & 4.1 & 8.6 & 12.7\\
HD 291       & 46 & -7 & -14 & 0.17 & 5.8 & 2.6 & 1.0 & 2.3 & 5.1 & 9.7 & 13.1 & 6.1 & 3.2 & 1.3 & 2.6 & 5.6 & 10.0 & 13.1 & 5.6 & 2.5 & 1.0 & 2.2 & 4.9 & 9.4 & 13.0\\
HD 307       & 12 & -59 & 0 & 0.23 & 9.0 & 10.4 & 3.0 & 5.4 & 9.1 & 12.4 & 13.7 & 8.4 & 9.3 & 2.4 & 4.7 & 8.6 & 12.1 & 13.7 & 8.8 & 9.2 & 2.9 & 5.3 & 8.9 & 12.3 & 13.7\\
HD 330       & -7 & -57 & -32 & 0.21 & 9.2 & 11.1 & 3.3 & 5.8 & 9.4 & 12.5 & 13.8 & 9.1 & 10.6 & 3.1 & 5.6 & 9.3 & 12.4 & 13.7 & 9.0 & 10.1 & 3.2 & 5.6 & 9.2 & 12.4 & 13.7\\
HD 334       & 5 & 5 & 30 & 0.07 & 7.6 & 6.0 & 2.4 & 4.2 & 7.3 & 11.1 & 13.4 & 7.7 & 6.2 & 2.6 & 4.4 & 7.4 & 11.1 & 13.4 & 7.8 & 6.7 & 2.5 & 4.4 & 7.6 & 11.3 & 13.5\\
HD 373       & -25 & -18 & 5 & 0.07 & 4.2 & 0.8 & 0.4 & 1.0 & 3.1 & 8.0 & 12.5 & 4.3 & 0.9 & 0.5 & 1.1 & 3.3 & 7.9 & 12.4 & 4.5 & 0.9 & 0.5 & 1.3 & 3.6 & 8.3 & 12.6\\
HD 392       & -26 & -9 & -1 & 0.05 & 3.3 & 0.5 & 0.2 & 0.6 & 2.1 & 6.5 & 11.8 & 3.3 & 0.6 & 0.3 & 0.6 & 2.1 & 6.5 & 11.8 & 3.1 & 0.6 & 0.3 & 0.7 & 2.0 & 5.9 & 11.4\\
HD 427       & -11 & 3 & 19 & 0.04 & 6.3 & 3.6 & 1.5 & 2.9 & 5.7 & 9.9 & 13.1 & 5.9 & 3.3 & 1.4 & 2.7 & 5.3 & 9.4 & 12.9 & 6.4 & 3.7 & 1.5 & 2.9 & 5.9 & 10.2 & 13.2\\
HD 439       & -44 & -57 & -7 & 0.23 & 8.4 & 9.9 & 2.1 & 4.5 & 8.5 & 12.2 & 13.7 & 8.4 & 9.4 & 2.3 & 4.7 & 8.5 & 12.1 & 13.7 & 8.1 & 8.0 & 1.9 & 4.3 & 8.2 & 12.0 & 13.7\\
HD 447       & -12 & 11 & -1 & 0.06 & 4.5 & 1.5 & 0.6 & 1.4 & 3.6 & 7.9 & 12.3 & 4.4 & 1.4 & 0.6 & 1.4 & 3.5 & 7.7 & 12.1 & 3.4 & 0.6 & 0.2 & 0.6 & 2.1 & 6.7 & 11.9\\
HD 457       & -31 & -17 & 14 & 0.07 & 5.9 & 2.6 & 1.1 & 2.3 & 5.2 & 9.8 & 13.1 & 5.8 & 2.6 & 1.1 & 2.3 & 5.2 & 9.7 & 13.0 & 5.5 & 2.5 & 1.1 & 2.2 & 4.8 & 9.2 & 12.9\\
HD 466       & -21 & -18 & 13 & 0.06 & 5.6 & 2.1 & 0.9 & 2.0 & 4.8 & 9.5 & 13 & 5.1 & 1.9 & 0.9 & 1.8 & 4.3 & 8.9 & 12.8 & 5.7 & 2.4 & 1.0 & 2.2 & 5.0 & 9.5 & 13.0\\
HD 547       & -22 & -63 & -12 & 0.24 & 9.0 & 11.2 & 2.8 & 5.4 & 9.2 & 12.4 & 13.8 & 8.8 & 10.1 & 2.7 & 5.2 & 9.0 & 12.3 & 13.7 & 8.8 & 9.6 & 2.7 & 5.2 & 9.0 & 12.4 & 13.7\\
HD 564       & 27 & 4 & -4 & 0.11 & 4.1 & 0.8 & 0.3 & 1.0 & 3.0 & 7.6 & 12.2 & 4.4 & 0.8 & 0.5 & 1.3 & 3.6 & 8.0 & 12.3 & 3.8 & 0.8 & 0.4 & 0.9 & 2.7 & 7.1 & 12.1\\
HD 578       & 26 & -3 & 18 & 0.11 & 6.5 & 3.8 & 1.5 & 3.0 & 6.0 & 10.3 & 13.2 & 6.7 & 4.2 & 1.7 & 3.2 & 6.3 & 10.4 & 13.2 & 6.3 & 3.6 & 1.5 & 2.9 & 5.8 & 10.0 & 13.2\\
HD 604       & -18 & -38 & -15 & 0.13 & 5.9 & 1.8 & 0.7 & 1.9 & 5.2 & 10.1 & 13.2 & 6.0 & 2.0 & 0.9 & 2.2 & 5.4 & 10.2 & 13.3 & 5.5 & 1.7 & 0.7 & 1.8 & 4.8 & 9.7 & 13.1\\
HD 610       & 1 & -11 & -6 & 0.04 & 1.8 & 0.1 & 0.0 & 0.1 & 0.6 & 3.7 & 10.2 & 1.0 & 0.2 & 0.0 & 0.1 & 0.3 & 1.3 & 7.6 & 0.9 & 0.2 & 0.0 & 0.1 & 0.3 & 1.3 & 6.8\\
HD 615       & 34 & -5 & -8 & 0.13 & 4.6 & 1.3 & 0.5 & 1.3 & 3.6 & 8.3 & 12.6 & 5.0 & 1.7 & 0.6 & 1.6 & 4.1 & 8.7 & 12.7 & 4.4 & 1.2 & 0.5 & 1.3 & 3.4 & 7.9 & 12.4\\
HD 631       & -14 & 11 & -30 & 0.07 & 6.2 & 3.6 & 1.5 & 2.8 & 5.6 & 9.8 & 13.0 & 5.7 & 3.2 & 1.3 & 2.6 & 5.1 & 9.1 & 12.7 & 6.1 & 3.2 & 1.3 & 2.6 & 5.6 & 10.0 & 13.2\\
HD 633       & -1 & 13 & -12 & 0.08 & 4.4 & 1.3 & 0.5 & 1.3 & 3.4 & 7.8 & 12.2 & 4.1 & 1.3 & 0.6 & 1.3 & 3.2 & 7.1 & 11.7 & 3.9 & 0.7 & 0.4 & 0.9 & 2.8 & 7.6 & 12.3\\
HD 700       & 3 & 11 & 5 & 0.08 & 4.8 & 1.6 & 0.7 & 1.6 & 3.9 & 8.3 & 12.5 & 4.8 & 2.0 & 0.8 & 1.8 & 4.0 & 8.1 & 12.3 & 4.8 & 1.4 & 0.6 & 1.5 & 3.9 & 8.7 & 12.7\\
HD 732       & -26 & -14 & -2 & 0.06 & 2.9 & 0.3 & 0.1 & 0.4 & 1.6 & 5.9 & 11.6 & 3.2 & 0.6 & 0.2 & 0.6 & 2.0 & 6.5 & 11.8 & 3.2 & 0.6 & 0.2 & 0.7 & 2.1 & 6.3 & 11.6\\
HD 804       & -33 & -63 & 20 & 0.24 & 9.7 & 13.0 & 3.8 & 6.5 & 10.0 & 12.8 & 13.8 & 9.5 & 12.5 & 3.7 & 6.3 & 9.9 & 12.7 & 13.8 & 9.5 & 12.4 & 3.6 & 6.3 & 9.8 & 12.7 & 13.8\\
HD 852       & -29 & -17 & -6 & 0.07 & 2.6 & 0.1 & 0.0 & 0.2 & 1.2 & 5.5 & 11.5 & 2.4 & 0.2 & 0.1 & 0.2 & 1.1 & 5.2 & 11.2 & 3.0 & 0.2 & 0.1 & 0.4 & 1.8 & 5.9 & 11.4\\
HD 900       & -8 & -3 & -19 & 0.01 & 3.9 & 0.8 & 0.4 & 0.9 & 2.7 & 7.2 & 12.1 & 3.5 & 0.7 & 0.4 & 0.8 & 2.4 & 6.5 & 11.6 & 3.3 & 0.7 & 0.4 & 0.8 & 2.2 & 6.2 & 11.5\\
HD 903       & -22 & -9 & 7 & 0.04 & 4.4 & 1.2 & 0.5 & 1.2 & 3.4 & 8.0 & 12.5 & 4.4 & 1.1 & 0.5 & 1.3 & 3.5 & 8.0 & 12.4 & 4.1 & 1.2 & 0.5 & 1.2 & 3.1 & 7.4 & 12.1\\
HD 912       & -38 & -10 & -1 & 0.08 & 4.2 & 0.9 & 0.4 & 1.0 & 3.1 & 7.8 & 12.4 & 3.9 & 0.7 & 0.4 & 0.9 & 2.8 & 7.4 & 12.2 & 4.1 & 1.0 & 0.5 & 1.1 & 3.1 & 7.5 & 12.2\\
HD 949       & -32 & -12 & 4 & 0.07 & 4.3 & 1.0 & 0.4 & 1.1 & 3.3 & 8.0 & 12.5 & 4.3 & 0.9 & 0.5 & 1.1 & 3.3 & 8.1 & 12.5 & 4.1 & 1.1 & 0.5 & 1.2 & 3.1 & 7.5 & 12.2\\
HD 975       & -4 & 3 & 1 & 0.04 & 3.6 & 0.7 & 0.3 & 0.8 & 2.5 & 6.8 & 11.9 & 3.4 & 0.7 & 0.4 & 0.9 & 2.4 & 6.3 & 11.3 & 3.4 & 0.6 & 0.3 & 0.7 & 2.2 & 6.5 & 11.8\\
HD 984       & -13 & -24 & -6 & 0.08 & 2.9 & 0.2 & 0.1 & 0.3 & 1.5 & 6.1 & 11.8 & 2.4 & 0.2 & 0.1 & 0.3 & 1.2 & 5.0 & 11.1 & 3.3 & 0.3 & 0.1 & 0.4 & 2.0 & 6.7 & 12.0\\
HD 1000      & 46 & -46 & 10 & 0.23 & 9.0 & 9.6 & 3.3 & 5.6 & 9.0 & 12.3 & 13.7 & 8.7 & 10.4 & 2.5 & 5.1 & 8.9 & 12.3 & 13.7 & 8.6 & 8.2 & 3.0 & 5.2 & 8.7 & 12.1 & 13.7\\
HD 1004      & 9 & -11 & -12 & 0.06 & 3.0 & 0.3 & 0.1 & 0.4 & 1.7 & 5.9 & 11.6 & 3.3 & 0.6 & 0.3 & 0.6 & 2.0 & 6.6 & 11.9 & 2.6 & 0.6 & 0.2 & 0.5 & 1.4 & 5.0 & 10.8\\
HD 1015      & 46 & -24 & -8 & 0.18 & 6.8 & 4.0 & 1.6 & 3.1 & 6.4 & 10.7 & 13.4 & 6.4 & 3.5 & 1.2 & 2.7 & 6.0 & 10.5 & 13.3 & 6.0 & 3.0 & 1.2 & 2.5 & 5.4 & 9.9 & 13.1\\
HD 1047      & 0 & 11 & 0 & 0.07 & 4.3 & 1.2 & 0.5 & 1.2 & 3.3 & 7.7 & 12.2 & 4.0 & 1.2 & 0.5 & 1.3 & 3.1 & 7.0 & 11.7 & 3.9 & 0.7 & 0.4 & 0.9 & 2.8 & 7.5 & 12.3\\
HD 1094      & -13 & -15 & -44 & 0.03 & 7.7 & 6.1 & 2.3 & 4.2 & 7.5 & 11.3 & 13.5 & 8.0 & 7.3 & 2.7 & 4.7 & 7.9 & 11.5 & 13.5 & 7.5 & 5.7 & 2.4 & 4.2 & 7.3 & 11.1 & 13.4\\
HD 1101      & -12 & -23 & -6 & 0.07 & 2.7 & 0.2 & 0.0 & 0.2 & 1.3 & 5.7 & 11.7 & 2.7 & 0.2 & 0.1 & 0.3 & 1.4 & 5.7 & 11.5 & 2.5 & 0.2 & 0.1 & 0.2 & 1.1 & 5.4 & 11.3\\
HD 1184      & 42 & -17 & -41 & 0.16 & 8.3 & 7.3 & 2.8 & 4.8 & 8.1 & 11.7 & 13.6 & 8.2 & 7.5 & 2.8 & 4.8 & 8.2 & 11.7 & 13.6 & 7.8 & 6.8 & 2.6 & 4.4 & 7.6 & 11.4 & 13.5\\
HD 1204      & 35 & -42 & -23 & 0.2 & 8.3 & 7.6 & 2.6 & 4.7 & 8.2 & 11.9 & 13.6 & 8.0 & 8.0 & 2.0 & 4.2 & 8.1 & 11.9 & 13.6 & 8.0 & 6.9 & 2.4 & 4.5 & 7.9 & 11.7 & 13.6\\
HD 1213      & -24 & -12 & 3 & 0.05 & 3.7 & 0.6 & 0.3 & 0.8 & 2.6 & 7.2 & 12.2 & 3.8 & 0.7 & 0.4 & 0.8 & 2.7 & 7.2 & 12.1 & 3.6 & 0.7 & 0.4 & 0.9 & 2.6 & 6.8 & 11.8\\
HD 1271      & -45 & -53 & 38 & 0.21 & 10.2 & 13.4 & 4.6 & 7.3 & 10.5 & 12.9 & 13.8 & 10.0 & 12.5 & 4.5 & 7.1 & 10.3 & 12.8 & 13.8 & 9.9 & 11.6 & 4.4 & 7.0 & 10.2 & 12.8 & 13.8\\
HD 1343      & -4 & -3 & 3 & 0.02 & 3.5 & 0.6 & 0.3 & 0.7 & 2.4 & 6.7 & 11.9 & 3.4 & 0.7 & 0.4 & 0.8 & 2.4 & 6.4 & 11.5 & 3.1 & 0.6 & 0.3 & 0.7 & 1.9 & 5.9 & 11.3\\
HD 1352      & -40 & -21 & -16 & 0.11 & 4.6 & 0.9 & 0.4 & 1.1 & 3.5 & 8.5 & 12.7 & 4.9 & 1.2 & 0.5 & 1.4 & 4.0 & 9.0 & 12.9 & 4.8 & 1.5 & 0.5 & 1.4 & 4.0 & 8.7 & 12.7\\
HD 1391      & 12 & 1 & -8 & 0.07 & 2.8 & 0.3 & 0.1 & 0.4 & 1.6 & 5.7 & 11.3 & 3.5 & 0.6 & 0.2 & 0.7 & 2.4 & 6.7 & 11.8 & 2.4 & 0.3 & 0.1 & 0.4 & 1.3 & 4.8 & 10.7\\
HD 1418      & 13 & -35 & -42 & 0.13 & 8.6 & 8.3 & 2.9 & 5.1 & 8.5 & 12.0 & 13.7 & 8.1 & 7.0 & 2.6 & 4.6 & 8.0 & 11.7 & 13.6 & 8.2 & 6.9 & 2.8 & 4.8 & 8.2 & 11.8 & 13.6\\
HD 1455      & 19 & -15 & 3 & 0.09 & 4.7 & 1.3 & 0.6 & 1.4 & 3.7 & 8.5 & 12.7 & 4.6 & 1.1 & 0.5 & 1.3 & 3.6 & 8.5 & 12.7 & 3.9 & 0.9 & 0.4 & 1.0 & 2.9 & 7.2 & 12.1\\
HD 1497      & 14 & -27 & -17 & 0.11 & 5.6 & 2.0 & 0.8 & 1.9 & 4.8 & 9.6 & 13.1 & 5.0 & 1.3 & 0.6 & 1.5 & 4.2 & 9.1 & 12.9 & 5.3 & 1.9 & 0.8 & 1.8 & 4.5 & 9.2 & 12.9\\
HD 1513      & 3 & -2 & -2 & 0.04 & 2.7 & 0.3 & 0.1 & 0.3 & 1.4 & 5.4 & 11.2 & 3.2 & 0.6 & 0.2 & 0.6 & 2.1 & 6.3 & 11.6 & 2.2 & 0.3 & 0.1 & 0.3 & 1.1 & 4.4 & 10.4\\
HD 1557      & -47 & 6 & -32 & 0.12 & 7.1 & 5.0 & 2.0 & 3.7 & 6.7 & 10.6 & 13.3 & 6.7 & 4.4 & 1.8 & 3.3 & 6.3 & 10.3 & 13.2 & 6.7 & 4.1 & 1.7 & 3.2 & 6.2 & 10.4 & 13.2\\
HD 1591      & -36 & -21 & -9 & 0.1 & 3.5 & 0.3 & 0.1 & 0.5 & 2.1 & 7.0 & 12.2 & 4.1 & 0.6 & 0.3 & 0.8 & 2.9 & 8.0 & 12.5 & 4.0 & 0.6 & 0.3 & 0.8 & 2.9 & 7.7 & 12.4\\
HD 1603      & 24 & -7 & -2 & 0.1 & 4.0 & 0.8 & 0.3 & 1.0 & 2.9 & 7.6 & 12.3 & 4.3 & 0.9 & 0.5 & 1.2 & 3.4 & 8.0 & 12.4 & 3.7 & 0.7 & 0.4 & 0.9 & 2.6 & 7.0 & 12.0\\
HD 1674      & 54 & -7 & 12 & 0.19 & 7.1 & 4.9 & 1.9 & 3.6 & 6.8 & 10.9 & 13.4 & 7.3 & 5.5 & 2.0 & 3.8 & 7.0 & 10.9 & 13.4 & 7.0 & 4.2 & 1.8 & 3.5 & 6.6 & 10.8 & 13.4\\
HD 1683      & 29 & -5 & -2 & 0.12 & 4.4 & 1.1 & 0.4 & 1.2 & 3.3 & 8.0 & 12.5 & 4.8 & 1.4 & 0.6 & 1.5 & 4.0 & 8.6 & 12.6 & 4.1 & 1.1 & 0.5 & 1.2 & 3.1 & 7.6 & 12.3\\
HD 1686      & 11 & 4 & -6 & 0.07 & 3.1 & 0.4 & 0.1 & 0.5 & 1.9 & 6.1 & 11.5 & 3.7 & 0.6 & 0.3 & 0.8 & 2.7 & 7.0 & 11.9 & 2.6 & 0.5 & 0.1 & 0.5 & 1.5 & 5.1 & 10.9\\
HD 1689      & -22 & 16 & 3 & 0.09 & 5.6 & 2.8 & 1.1 & 2.3 & 4.9 & 9.1 & 12.8 & 4.7 & 2.0 & 0.9 & 1.8 & 3.9 & 7.9 & 12.1 & 4.5 & 0.7 & 0.4 & 1.1 & 3.5 & 8.4 & 12.7\\
HD 1764      & -1 & -41 & 8 & 0.15 & 7.1 & 4.5 & 1.5 & 3.3 & 6.8 & 11.2 & 13.5 & 6.6 & 3.4 & 1.3 & 2.8 & 6.2 & 10.7 & 13.4 & 7.0 & 4.5 & 1.6 & 3.3 & 6.7 & 11.0 & 13.4\\
HD 1778      & 50 & -16 & -23 & 0.18 & 7.1 & 4.6 & 1.8 & 3.4 & 6.7 & 10.9 & 13.4 & 7.0 & 4.2 & 1.7 & 3.4 & 6.7 & 10.8 & 13.4 & 6.5 & 3.9 & 1.5 & 3.0 & 6.0 & 10.3 & 13.2\\
HD 1828      & -10 & -4 & -24 & 0.01 & 4.7 & 1.5 & 0.7 & 1.5 & 3.8 & 8.3 & 12.6 & 4.2 & 1.2 & 0.6 & 1.3 & 3.3 & 7.5 & 12.1 & 4.1 & 1.2 & 0.6 & 1.3 & 3.2 & 7.4 & 12.1\\
HD 1856      & -25 & -28 & -15 & 0.1 & 4.3 & 0.6 & 0.3 & 0.9 & 3.1 & 8.3 & 12.7 & 4.0 & 0.7 & 0.4 & 0.9 & 2.9 & 7.7 & 12.4 & 4.4 & 0.7 & 0.4 & 1.0 & 3.4 & 8.4 & 12.7\\
HD 1898      & 33 & 6 & -8 & 0.14 & 4.5 & 1.2 & 0.5 & 1.2 & 3.5 & 8.1 & 12.5 & 5.1 & 1.6 & 0.7 & 1.7 & 4.4 & 8.8 & 12.6 & 4.9 & 1.8 & 0.7 & 1.6 & 4.0 & 8.6 & 12.7\\
HD 1952      & 14 & -8 & 2 & 0.07 & 3.8 & 0.7 & 0.3 & 0.9 & 2.7 & 7.3 & 12.2 & 4.0 & 0.8 & 0.4 & 1.0 & 3.0 & 7.6 & 12.3 & 3.5 & 0.7 & 0.4 & 0.8 & 2.4 & 6.6 & 11.8\\
HD 1980      & -34 & -7 & 6 & 0.07 & 4.9 & 1.6 & 0.7 & 1.6 & 3.9 & 8.6 & 12.7 & 4.6 & 1.3 & 0.6 & 1.4 & 3.7 & 8.3 & 12.6 & 4.5 & 1.4 & 0.6 & 1.5 & 3.6 & 8.1 & 12.4\\
HD 2021      & -15 & -4 & -12 & 0.02 & 2.8 & 0.3 & 0.1 & 0.4 & 1.6 & 5.6 & 11.3 & 2.7 & 0.4 & 0.1 & 0.5 & 1.6 & 5.3 & 10.9 & 2.3 & 0.3 & 0.1 & 0.4 & 1.2 & 4.5 & 10.4\\
HD 2121      & -53 & -30 & 9 & 0.16 & 6.6 & 3.3 & 1.2 & 2.7 & 6.2 & 10.8 & 13.4 & 6.9 & 4.1 & 1.4 & 3.1 & 6.6 & 11 & 13.4 & 6.7 & 4.1 & 1.3 & 3.0 & 6.4 & 10.7 & 13.4\\
HD 2302      & -29 & -23 & -8 & 0.09 & 2.9 & 0.2 & 0.0 & 0.3 & 1.5 & 6.2 & 11.9 & 3.1 & 0.2 & 0.1 & 0.3 & 1.8 & 6.5 & 11.9 & 3.1 & 0.2 & 0.1 & 0.3 & 1.8 & 6.6 & 12.0\\
HD 2305      & -3 & -16 & -10 & 0.05 & 2.2 & 0.1 & 0.0 & 0.1 & 0.9 & 4.5 & 10.9 & 0.8 & 0.2 & 0.0 & 0.1 & 0.3 & 1.1 & 6.6 & 2.1 & 0.2 & 0.1 & 0.2 & 0.9 & 4.6 & 10.8\\
HD 2330      & -6 & -25 & -12 & 0.08 & 3.7 & 0.5 & 0.2 & 0.7 & 2.5 & 7.4 & 12.4 & 3.3 & 0.6 & 0.3 & 0.7 & 2.1 & 6.7 & 12.0 & 3.9 & 0.6 & 0.4 & 0.8 & 2.7 & 7.5 & 12.3\\[+0.10cm]
		\hline
	\end{tabular}
\end{table*}

In this Appendix, we present the results obtained by the three kinematical methods for the first 75 stars of the sample. The full data is only available in electronic format. The superscripts 1, 2 and 3 denotes the kinematical method used, respectively Method $UVW$, Method $eVW$ and Method $eUW$. The columns descriptions are as follows:

\begin{itemize}
\item[Name:] as given in the original Geneva-Copenhagen Catalogue.
\item[$U$:] GCSIII heliocentric space velocity $U$ ($\kms$).
\item[$V$:] GCSIII heliocentric space velocity $V$ ($\kms$).
\item[$W$:] GCSIII heliocentric space velocity $W$ ($\kms$).
\item[$e$:] GCSIII eccentricity of galactic orbit.
\item[$t_{\mathrm{E}}$:] expected age (Gyr).
\item[$t_{\mathrm{ML}}$:] most likely age (Gyr).
\item[$t_{2.5}$:] 2.5\% percentile from age pdf (Gyr).
\item[$t_{16}$:] 16\% percentile (Gyr).
\item[$t_{50}$:] 50\% percentile (Gyr).
\item[$t_{84}$:] 84\% percentile (Gyr).
\item[$t_{97.5}$:] 97.5\% percentile (Gyr).
\end{itemize}


\bsp	
\label{lastpage}
\end{document}